\documentclass[aps,pra,reprint,twocolumn,superscriptaddress,floatfix]{revtex4-2}
\pdfoutput=1

\usepackage[utf8]{inputenc}
\usepackage[english]{babel}
\usepackage[T1]{fontenc}
\usepackage{siunitx}
\usepackage{amsmath}
\usepackage{braket}
\usepackage{tikz}
\usepackage{hyperref}
\usepackage[capitalize]{cleveref}

\newcommand{\NKD}[3]{[[#1,#2,#3]]}

\begin{document}

\title{Implementing fault-tolerant non-Clifford gates using the [[8,3,2]] color code}

\author{Daniel Honciuc Menendez}
\affiliation{Department of Physics, University of Toronto, Toronto, ON M5S 1A7, Canada}
\author{Annie Ray}
\affiliation{Institute for Quantum Computing, University of Waterloo, Waterloo, ON N2L 3G1, Canada}
\affiliation{Perimeter Institute for Theoretical Physics, Waterloo, ON N2L 2Y5, Canada}
\author{Michael Vasmer}
\affiliation{Institute for Quantum Computing, University of Waterloo, Waterloo, ON N2L 3G1, Canada}
\affiliation{Perimeter Institute for Theoretical Physics, Waterloo, ON N2L 2Y5, Canada}

\begin{abstract}
    Quantum computers promise to solve problems that are intractable for classical computers, but qubits are vulnerable to many sources of error, limiting the depth of the circuits that can be reliably executed on today's quantum hardware.
    Quantum error correction has been proposed as a solution to this problem, whereby quantum information is protected by encoding it into a quantum error-correcting code.
    But protecting quantum information is not enough, we must also process the information using logic gates that are robust to faults that occur during their execution.
    One method for processing information fault-tolerantly is to use quantum error-correcting codes that have logical gates with a tensor product structure (transversal gates), making them naturally fault-tolerant.
    Here, we test the performance of a code with such transversal gates, the \NKD{8}{3}{2} color code, using trapped-ion and superconducting hardware.
    We observe improved performance (compared to no encoding) for encoded circuits implementing non-Clifford gates, a class of gates that are essential for achieving universal quantum computing.
    In particular, we find improved performance for an encoded circuit implementing the control-control $Z$ gate, a key gate in Shor's algorithm.
    Our results illustrate the potential of using codes with transversal gates to implement non-trivial algorithms on near-term quantum hardware.
\end{abstract}

\date{August 1, 2025}
\maketitle

\section{Introduction \label{sec:intro}}

Quantum error correction (QEC) promises to unlock the full potential of quantum computing, by protecting fragile qubits from the effects of decoherence~\cite{shorSchemeReducingDecoherence1995,Steane1996QEC,chuang1995QuantumErrorCorrection}.
But it is not enough to merely preserve the quantum information stored in a qubit register, we also need to perform a universal set of logical gates in a fault-tolerant manner~\cite{shorFaulttolerantQuantumComputation1996}.
Logical gates in the Clifford group (the unitaries that map Pauli operators to Pauli operators) are often relatively straightforward to implement fault-tolerantly in a given QEC code, however they are not universal.
In fact, no QEC code can have a transversal and universal set of logical gates~\cite{eastinRestrictionsTransversalEncoded2009}.
To obtain a universal gate set we need an additional non-Clifford gate~\cite{nebeInvariantsCliffordGroups2001}, but implementing gates from this class fault-tolerantly is often difficult, usually requiring complex procedures such as magic state distillation~\cite{bravyiUniversalQuantumComputation2005,knillFaultTolerantPostselectedQuantum2004a}.

Certain QEC codes with special structure have transversal non-Clifford gates, where a transversal gate is a gate that acts as a tensor product unitaries that do not entangle different qubits in the same QEC code block.
Examples of such gates include the transversal CNOT available in all CSS codes, and any gate acting as a tensor product of single-qubit unitaries.
Transversal gates are naturally fault-tolerant as they do not spread errors within a code block.

There exists a family of codes known as triorthogonal codes~\cite{bravyiMagicstateDistillationLow2012} with transversal non-Clifford gates, implemented by tensor products of $T = \mathrm{diag}\left(1,\exp(i\pi/4)\right)$
Certain (generalized) triorthogonal codes have transversal entangling non-Clifford gates, the smallest of which (to our knowledge) is the \NKD{8}{3}{2} color code~\cite{kubicaUnfoldingColorCode2015,campbell2016blog}, which has a transversal $\mathrm{CCZ} = \mathrm{diag}(1,1,1,1,1,1,1,-1)$ gate.
From a fault-tolerance perspective, it is particularly desirable to implement complex entangling gates using single-qubit gates, as single-qubit gates are often an order of magnitude less noisy than entangling gates in many hardware platforms~\cite{Ballance2016HighFidelity,wright2019Benchmarking11qubitQuantum,jurcevic2021DemonstrationQuantumVolume,Wu2021QuantumAdvantage,huang2019FidelityBenchmarksTwoqubit,moses2023RaceTrackTrappedIon}.
Using small codes to demonstrate fault-tolerant Clifford and non-Clifford operations has previously been suggested~\cite{gottesman_quantum_2016} and implemented in NMR~\cite{souza2011ExperimentalMagicState,Zhang2012Experimental}, trapped-ion~\cite{niggExperimentalQuantumComputations2014,eganFaultTolerantOperationQuantum2021,postler2022DemonstrationFaulttolerantUniversal,ryan-anderson2022ImplementingFaulttolerantEntangling}, and superconducting hardware~\cite{vuillotErrorDetectionHelpful2018,harperFaultTolerantLogicalGates2019,gupta2023EncodingMagicState}. 

Here, we investigate the performance of the encoded gates of the \NKD{8}{3}{2} code on superconducting and trapped-ion hardware platforms.
We compare the performance of the encoded gates with the same gates executed with no encoding, finding that the encoded gates perform better than their bare (non-encoded) counterparts in every case where the encoded gate is non-Clifford, even though the encoded circuits contain more entangling gates than the bare circuits.
Notably, we observe improved performance for the CCZ gate, which is the dominant gate in circuits such as adders~\cite{Vedral1996QArithmetic,Gidney2018halvingcostof} and the modular exponentiation used in Shor's algorithm~\cite{shorPolynomialTimeAlgorithmsPrime1997,Gidney2021howtofactorbit}.

The remainder of this article is structured as follows.
In \cref{sec:832}, we review the definition of the \NKD{8}{3}{2} code and its transversal logical gates.
In \cref{sec:ft}, we give fault-tolerant circuits for preparing encoded states of the \NKD{8}{3}{2} code and for logical measurements.
In \cref{sec:results}, we describe our demonstrations on quantum hardware and their results, and we conclude with \cref{sec:discuss}.

\section{The \texorpdfstring{\NKD{8}{3}{2}}{[[8,3,2]]} color code}
\label{sec:832}

\begin{figure}[ht]
    \centering
    (a)
    \begin{tikzpicture}[scale=0.3]
        \coordinate[label=below:7] (A) at (0,0,0);
        \coordinate[label=below:3] (B) at (0,0,4);
        \coordinate[label=above:6] (C) at (0,4,0);
        \coordinate[label=above:2] (D) at (0,4,4);
        \coordinate[label=below:5] (E) at (4,0,0);
        \coordinate[label=below:1] (F) at (4,0,4);
        \coordinate[label=above:4] (G) at (4,4,0);
        \coordinate[label=above:0] (H) at (4,4,4);
        \draw [dashed] (A) -- (B);
        \draw [dashed] (A) -- (C);
        \draw [dashed] (A) -- (E);
        \draw (B) -- (D);
        \draw (B) -- (F);
        \draw (C) -- (D);
        \draw (C) -- (G);
        \draw (D) -- (H);
        \draw (E) -- (F);
        \draw (E) -- (G);
        \draw (F) -- (H);
        \draw (G) -- (H);
    \end{tikzpicture}
    \quad
    (b)
    \begin{tikzpicture}[scale=0.3]
        \coordinate[label=below:7] (A) at (0,0,0);
        \coordinate[label=below:3] (B) at (0,0,4);
        \coordinate[label=above:6] (C) at (0,4,0);
        \coordinate[label=above:2] (D) at (0,4,4);
        \coordinate[label=below:5] (E) at (4,0,0);
        \coordinate[label=below:1] (F) at (4,0,4);
        \coordinate[label=above:4] (G) at (4,4,0);
        \coordinate[label=above:0] (H) at (4,4,4);
        \draw [dashed] (A) -- (B);
        \draw [dashed] (A) -- (C);
        \draw [dashed] (A) -- (E);
        \draw (B) -- (D);
        \draw (B) -- (F);
        \draw (C) -- (D);
        \draw (C) -- (G);
        \draw (D) -- (H);
        \draw (E) -- (F);
        \draw (E) -- (G);
        \draw (F) -- (H);
        \draw (G) -- (H);
        \draw[fill=blue, fill opacity=0.2] (B) -- (F) -- (H) -- (D) -- cycle;
    \end{tikzpicture}
    \quad
    (c)
    \begin{tikzpicture}[scale=0.3]
        \coordinate[label=below:7] (A) at (0,0,0);
        \coordinate[label=below:3] (B) at (0,0,4);
        \coordinate[label=above:6] (C) at (0,4,0);
        \coordinate[label=above:2] (D) at (0,4,4);
        \coordinate[label=below:5] (E) at (4,0,0);
        \coordinate[label=below:1] (F) at (4,0,4);
        \coordinate[label=above:4] (G) at (4,4,0);
        \coordinate[label=above:0] (H) at (4,4,4);
        \draw [dashed] (A) -- (B);
        \draw [dashed] (A) -- (C);
        \draw [dashed] (A) -- (E);
        \draw (B) -- (D);
        \draw (B) -- (F);
        \draw (C) -- (D);
        \draw (C) -- (G);
        \draw (D) -- (H);
        \draw (E) -- (F);
        \draw (E) -- (G);
        \draw (F) -- (H);
        \draw (G) -- (H);
        \draw[fill=red, fill opacity=0.2] (E) -- (F) -- (H) -- (G) -- cycle;
        \draw[fill=red, fill opacity=0.2] (B) -- (F) -- (H) -- (D) -- cycle;
        \draw[fill=red, fill opacity=0.2] (C) -- (D) -- (H) -- (G) -- cycle;
    \end{tikzpicture}
    \caption{
    Geometric representation of the \NKD{8}{3}{2} code. 
    (a) The physical qubits reside at the vertices of the cube.
    (b) $Z$-type stabilizers are associated with faces, for example the blue face has an associated stabilizer $Z_0 Z_1 Z_2 Z_3$.
    (c) The $X$-type stabilizer acts on all the qubits.
    }
    \label{fig:832-cube}
\end{figure}
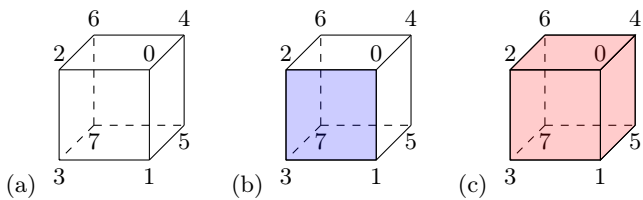

The \NKD{8}{3}{2} color code is a stabilizer code~\cite{gottesmanStabilizerCodesQuantum1997}, encoding 3 logical qubits into 8 physical qubits with distance 2 (meaning that it can detect any single-qubit error).
It is convenient to define the code using a geometric representation, where the physical qubits reside at the vertices of a cube, as shown in \cref{fig:832-cube}.
The stabilizer group is generated by an $X$-type operator acting on all the qubits, and by $Z$-type operators associated with the faces of the cube.
Concretely, using the qubit indices in \cref{fig:832-cube}, the stabilizer group is
\begin{equation}
\begin{split}
    \mathcal S = \langle 
    X^{\otimes 8}, 
    &Z_0 Z_1 Z_2 Z_3, 
    Z_4 Z_5 Z_6 Z_7, \\
    &Z_0 Z_1 Z_4 Z_5,
    Z_0 Z_2 Z_4 Z_6
    \rangle,
\end{split}
\label{eq:832-stab}
\end{equation}
where $Z_i$ denotes a Pauli $Z$ operator acting on qubit $i$ etc.
We note that the stabilizer generators in \cref{eq:832-stab} are either $X$-type or $Z$-type, meaning that the \NKD{8}{3}{2} code is a CSS code~\cite{calderbankGoodQuantumErrorcorrecting1996,steaneMultipleparticleInterferenceQuantum1996}.

The logical operators of the \NKD{8}{3}{2} code also have a geometric interpretation.
Logical $X$ operators are associated with the faces of the cube, and logical $Z$ operators with the edges of the cube.
We can choose the following basis of logical Pauli operators
\begin{equation}
\begin{split}
    \overline{X}_1 = X_0 X_1 X_2 X_3, \quad& \overline{Z}_1 = Z_0 Z_4, \\
    \overline{X}_2 = X_0 X_1 X_4 X_5, \quad& \overline{Z}_2 = Z_0 Z_2, \\
    \overline{X}_3 = X_0 X_2 X_4 X_6, \quad& \overline{Z}_3 = Z_0 Z_1,
\end{split}
\label{eq:logical-pauli}
\end{equation}
where we use overlines to distinguish operators acting on the logical qubits from operators acting on the physical qubits.

The \NKD{8}{3}{2} code is notable for having a non-Clifford transversal gate, $\mathrm{CCZ}$ implemented by $T$ and $T^\dagger$ gates.
Specifically, 
\begin{equation}
    \overline{\mathrm{CCZ}} = T_0 T_1^\dagger T_2^\dagger T_3 T_4^\dagger T_5 T_6 T_7^\dagger.
\end{equation}
This gate again has a geometric interpretation: vertices and edges of the cube form a bipartite graph and CCZ is implemented by applying $T$ to (the qubits on) one set of the vertices and $T^\dagger$ to the other.
The transversality of CCZ and Pauli $X$ imply that the \NKD{8}{3}{2} code also has transversal $\mathrm{CZ}=\mathrm{diag}(1,1,1,-1)$ gates, as follows
\begin{equation}
\begin{split}
    &\overline{\mathrm{CZ}}_{12} = S_0 S_2^\dagger S_4^\dagger S_6, \\
    &\overline{\mathrm{CZ}}_{13} = S_0 S_1^\dagger S_4^\dagger S_5, \\
    &\overline{\mathrm{CZ}}_{23} = S_0 S_1^\dagger S_2^\dagger S_3,
\end{split}
\end{equation}
where $S = T^2$ and $\mathrm{CZ}_{ij}$ acts on logical qubits $i$ and $j$.

\section{Fault-tolerant circuits}
\label{sec:ft}

For an error-detecting code such as the \NKD{8}{3}{2} code, we say that a circuit is fault-tolerant if any single-qubit error on the input state or an error at any single location in the circuit can at worst lead to a detectable error on the output state.
A circuit location can be a state preparation, gate, or measurement.
We need only consider Pauli errors due to error discretization~\cite{nielsenchuang}.
And we note that as the \NKD{8}{3}{2} code is a CSS code, it is sufficient to analyse $X$ and $Z$ errors independently.
We remark that the logical CCZ and CZ gates discussed in \cref{sec:832} are transversal and are therefore trivially fault-tolerant.
We also need fault-tolerant circuits for logical measurement and logical state preparation, and we now discuss each of these in turn.

As the \NKD{8}{3}{2} code is a CSS code, we can do a fault-tolerant measurement of the logical qubits in the $X$ or $Z$ basis by measuring all of the physical qubits in the $X$ or $Z$ basis, respectively, and processing the classical outcomes~\cite{nielsenchuang}.
In the case of an error-detecting code such as the \NKD{8}{3}{2} code, the classical processing is especially simple: we simply discard any measurement result that corresponds to a state that is not a $+1$ eigenvalue of the stabilizers.
For example, when measuring in the $X$ basis we accept any result whose parity is even, i.e., a $+1$ eigenstate of $X^{\otimes 8}$.
This is fault-tolerant because single-qubit errors before the measurements are detectable by definition, and any single measurement error is equivalent to a single-qubit error before the measurement.

\subsection{GHZ state preparation}

\begin{figure}[ht]
    \centering
    \includegraphics[width=.7\columnwidth]{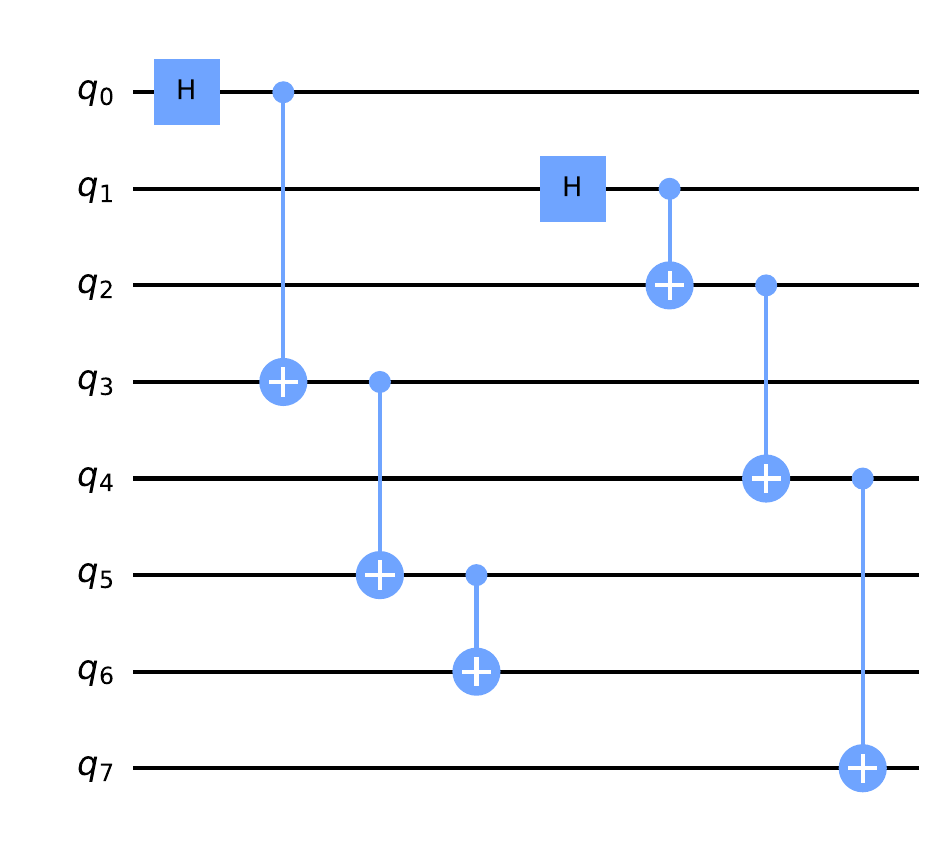}
    \caption{
        Fault-tolerant circuit for preparing the $\ket{\mathrm{GHZ}}$ state in the \NKD{8}{3}{2} code.
    }
    \label{fig:state-prep-ghz}
\end{figure}

\begin{figure*}[ht]
    \centering
    \includegraphics[width=.9\linewidth]{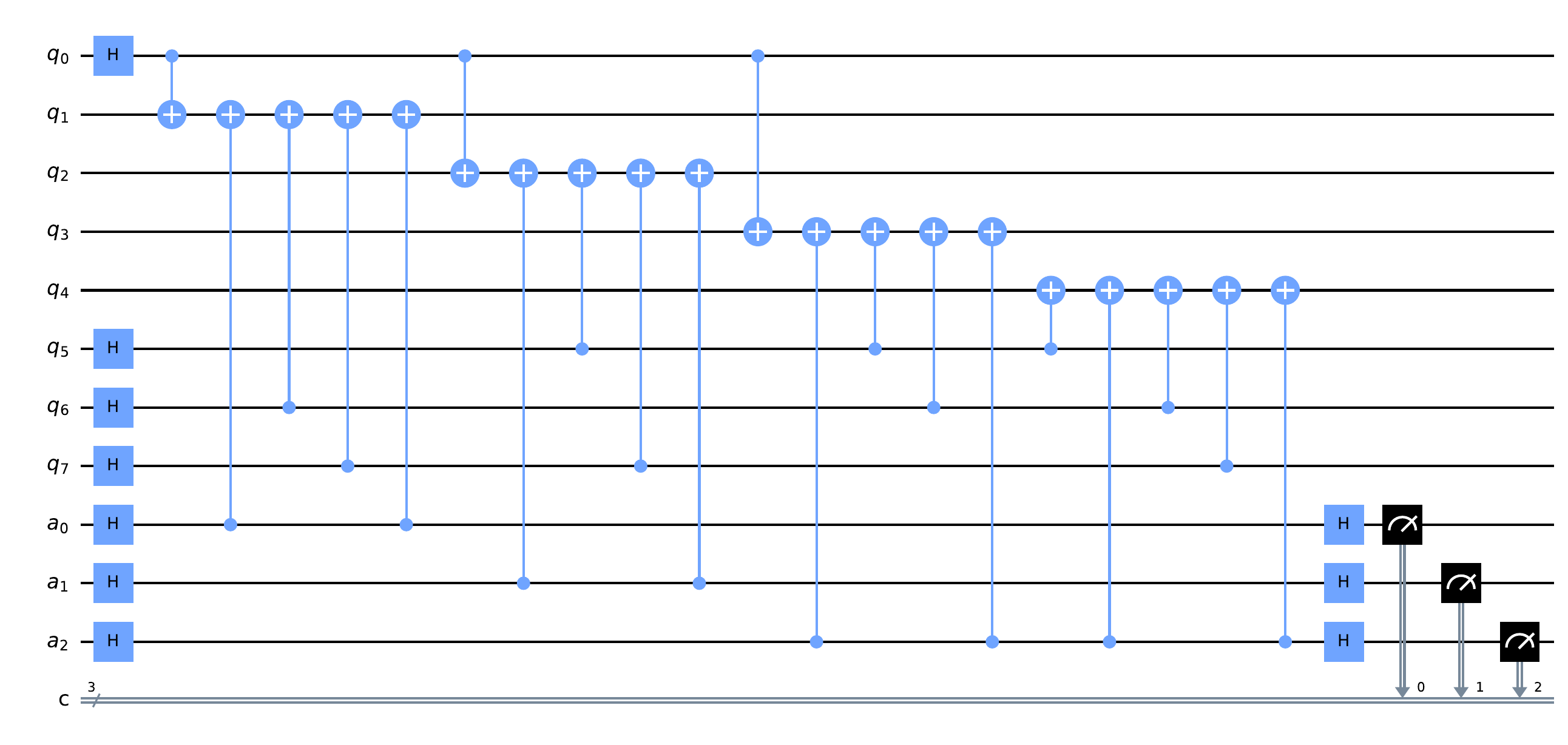}
    \caption{
        Fault-tolerant circuit for preparing the state $\ket{+++}$ in the \NKD{8}{3}{2} code. 
        The qubits $a_1$, $a_2$ and $a_3$ are flag qubits whose purpose is to detect certain $Z$ errors that could cause logical errors.
        If we measure the three flag qubits to be in the $\ket{0}$ state then we accept the output.
    }
    \label{fig:state-prep-+++}
\end{figure*}

First we consider a fault-tolerant circuit for preparing the logical GHZ state, $\ket{\mathrm{GHZ}} = (\ket{000} + \ket{111})/\sqrt 2$. 
Our circuit (shown in \cref{fig:state-prep-ghz}) factorizes into two independent and identical sub-circuits acting on qubits 0, 3, 5, 6 and qubits 1, 2, 4, 7 (the two bipartite sets discussed in \cref{sec:832}).
The \NKD{8}{3}{2} code can detect any weight $\leq 3$ $X$ error and so we only need to consider the four-qubit errors $X_0 X_3 X_5 X_6$ and $X_1 X_2 X_4 X_7$. 
However, each of these errors is in fact a logical $\overline X_1 \overline X_2 \overline X_3$ operator and so leaves the target $\ket{\mathrm{GHZ}}$ state invariant.
The only possible $Z$ errors are weight one (detectable) and weight two (non-detectable).
However, one can verify that all the non-detectable errors have trivial action on the target $\ket{\mathrm{GHZ}}$ state.
For example, the first CNOT could fail giving a $Z_1 Z_2$ error, but this implements a logical $\overline Z_2 \overline Z_3$ operator (see \cref{eq:logical-pauli}) and hence leaves the target $\ket{\mathrm{GHZ}}$ state invariant.

\subsection{\texorpdfstring{$\ket{+++}$}{|+++>} state preparation}

Next, we provide a fault-tolerant circuit for preparing the $\ket{+++}$ state, shown in \cref{fig:state-prep-+++}. 
In this circuit, the potentially problematic errors are those that can propagate through the CNOT gates.
Consider, for example, the CNOT gates with qubit 0 as the control.
The possible multi-qubit $X$ errors that can arise from these gates are
\begin{equation}
\begin{split}
    &X_0 X_3 \quad (\mathrm{detectable}), \\
    &X_0 X_2 X_3 \quad (\mathrm{detectable}), \\
    &X_0 X_1 X_2 X_3 \quad (\overline X_1),
\end{split}
\end{equation}
where the only non-detectable error has trivial action on the target encoded state. 
The same is true for the other groups of CNOT gates with the same target.
Certain $Z$ errors can also propagate through CNOT gates. 
For example, consider the CNOT gates with qubit 1 as the target.
The possible multi-qubit $Z$ errors that can arise from these gates are
\begin{equation}
\begin{split}
    &Z_1 Z_{a_0} \quad (\mathrm{detectable}), \\
    &Z_1 Z_7 Z_{a_0} \quad (\mathrm{detectable}), \\
    &Z_1 Z_6 Z_7 \quad (\mathrm{detectable}), \\
    &Z_1 Z_6 Z_7 Z_{a_0} \quad (\mathrm{detectable}), \\
    &Z_0 Z_1 Z_6 Z_7 \quad (\mathrm{stabilizer}).
\end{split}
\end{equation}
The purpose of the flag qubit~\cite{chao_quantum_2018}, $a_0$, is to make the error $Z_1 Z_7 = \overline Z_1 \overline Z_2$ detectable.
Similarly, the flag qubits $a_1$ and $a_2$ catch the errors $Z_2 Z_7$, $Z_3 Z_6$ and $Z_4 Z_6$.

\section{Demonstration results}
\label{sec:results}

We investigate the performance of circuits comprised of three parts: state preparation, a transversal logical gate, and logical measurement.

For the state preparation part, we consider either $\ket{\mathrm{GHZ}}$ or $\ket{+++}$ state preparation, using the circuits described in \cref{sec:ft}.
For the logical gate part, we consider each of the distinct products of the transversal logical CCZ, $\mathrm{CZ}_{12}$, $\mathrm{CZ}_{02}$ and $\mathrm{CZ}_{01}$ gates available in the \NKD{8}{3}{2} code, along with the logical identity gate implemented as a `no operation'.
For the logical measurement part, we consider transversal $Z$ basis and $X$ basis measurements.
In the encoded case, the fault-tolerant measurement involves post-selection and we provide the post-selection rates for each of the demonstrations in \cref{app:post}.

\begin{figure*}[ht]
\centering
    \includegraphics[width=.9\linewidth]{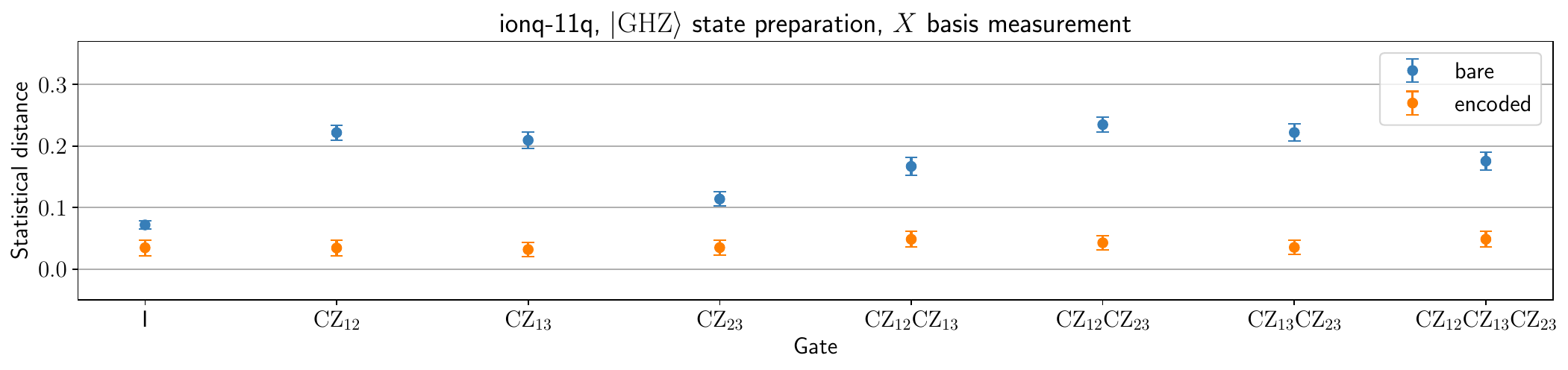}
    \includegraphics[width=.9\linewidth]{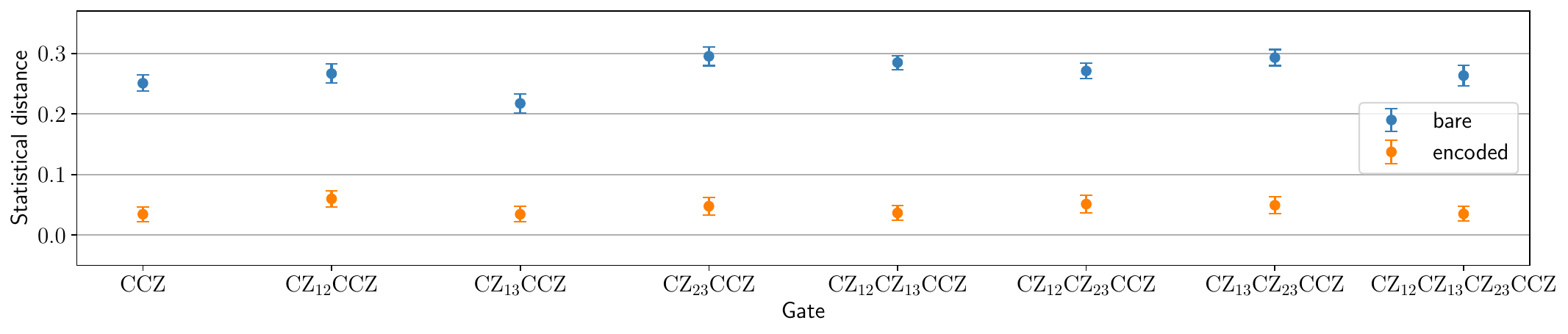}
    \includegraphics[width=.9\linewidth]{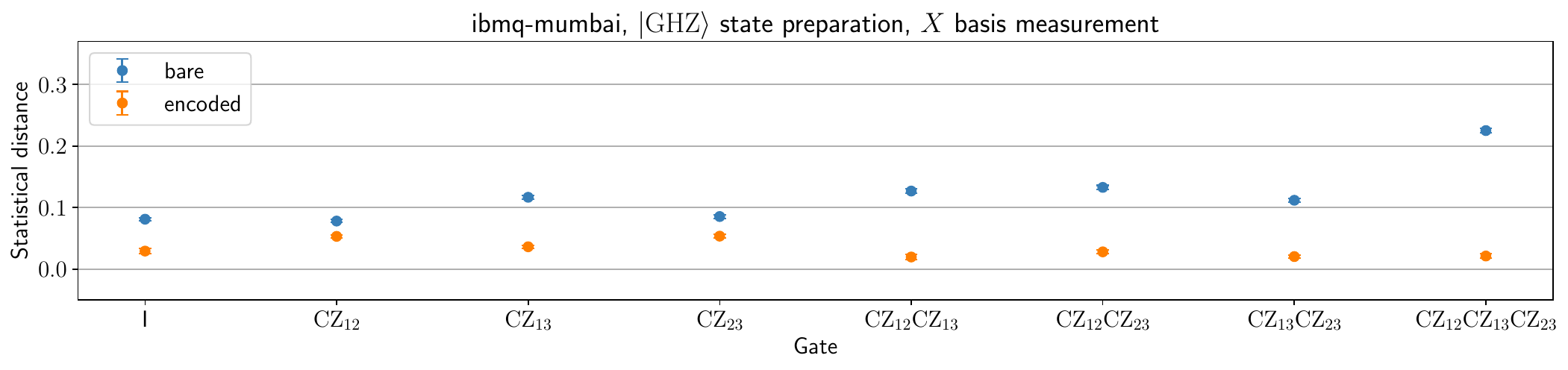}
    \includegraphics[width=.9\linewidth]{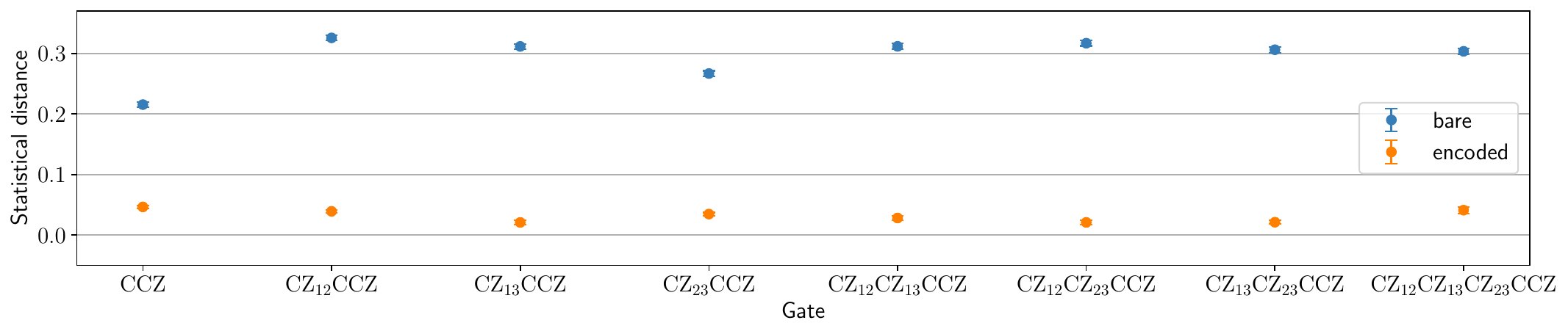}
    \caption{
    Performance of bare (non-encoded) and encoded versions of circuits for preparing states of the form $g \ket{\mathrm{GHZ}}$, where $g$ is a transversal gate of the \NKD{8}{3}{2} code.
    In each case, we measure the qubits in the $X$ basis and we plot the statistical distance of the observed measurement distribution from the ideal distribution.
    The upper two plots show the data for \textsf{ionq-11q}, where we ran 1024 shots for each circuit, and the lower two plots show the data for \textsf{ibmq\_mumbai} where we ran 10,000 shots for each circuit.
    In both cases, the error bars are calculated using bootstrap resampling.
    }
    \label{fig:X-GHZ}
\end{figure*}

We test these circuits on two quantum computers: \textsf{ibmq\_mumbai}, a 27-qubit device developed by IBM~\cite{ibmq}, and \textsf{ionq-11q}, an 11-qubit device developed by IonQ~\cite{wright2019Benchmarking11qubitQuantum}.
The IonQ device has all-to-all qubit connectivity, whereas the IBM device has ``heavy-hexagon'' qubit connectivity~\cite{hertzberg2021LaserannealingJosephsonJunctions}; see \cref{app:char} more details on the devices and their characteristics at the time of the demonstrations.
We only consider $\ket{\mathrm{GHZ}}$ state preparation on the IBM device,
as our circuit for preparing logical $\ket{+++}$ states (\cref{fig:state-prep-+++}) does not respect the connectivity constraints of the IBM device and would therefore require SWAP gates to implement, meaning that our error analysis is no longer valid. 
We leave open the possibility of finding a fault-tolerant circuit for preparing logical $\ket{+++}$ states on the IBM device.
We compare the performance of the encoded circuits against the performance of the bare (no encoding) circuits, using the statistical distance of the output distribution from the ideal output distribution as our metric.
The data and analysis code is available at~\cite{github}.

We show the results for $\ket{\mathrm{GHZ}}$ state preparation and $X$ basis measurement in \cref{fig:X-GHZ}.
For both devices and for every transversal gate, we observe improved performance of the encoded version of the circuit.
The results for $Z$ basis measurement are qualitatively similar; see \cref{app:Z-basis}.

\begin{figure*}[ht]
    \centering
    \includegraphics[width=.9\linewidth]{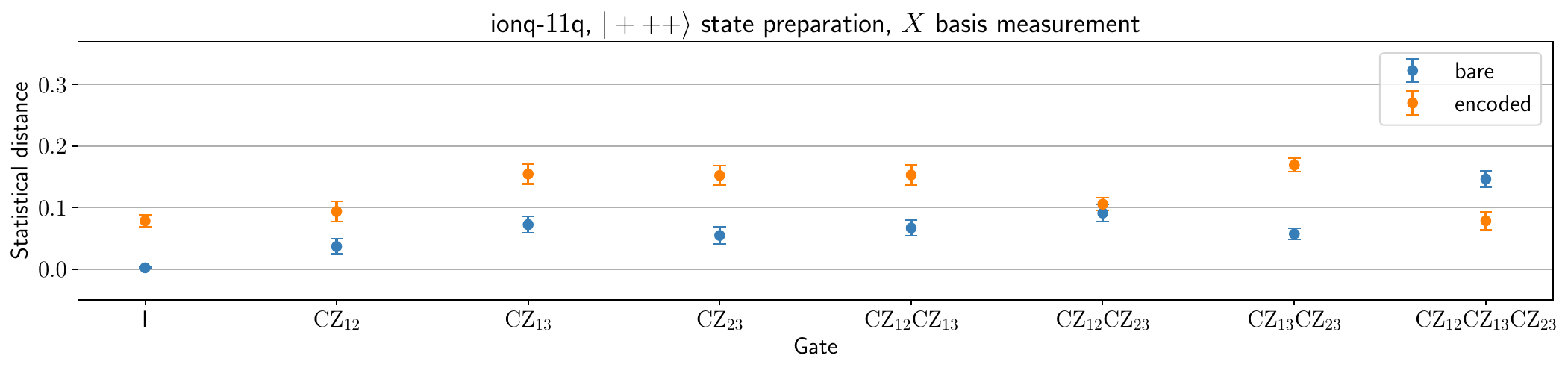}
    \includegraphics[width=.9\linewidth]{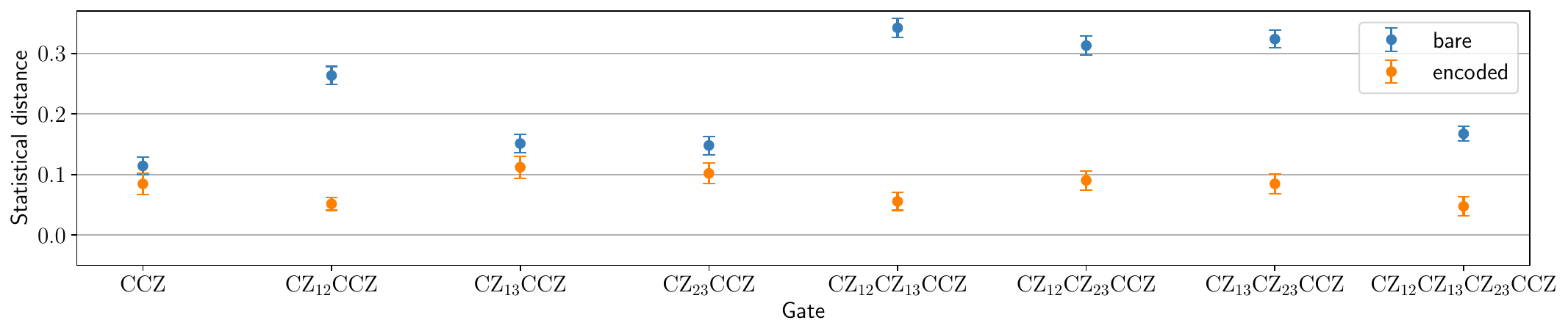}
    \caption{
    Performance of bare (non-encoded) and encoded versions of circuits for preparing states of the form $g \ket{+++}$, where $g$ is a transversal gate of the \NKD{8}{3}{2} code.
    In each case, we measure the qubits in the $X$ basis and we plot the statistical distance of the observed measurement distribution from the ideal distribution.
    Each data point represents 1024 shots of the circuit performed on \textsf{ionq-11q}, and we use bootstrap resampling to calculate the error bars.
    }
    \label{fig:X-+++}
\end{figure*}

We show the results for $\ket{\mathrm{+++}}$ state preparation and $X$ basis measurement in \cref{fig:X-+++}.
The bare version of the circuit performs better for transversal Clifford gates, whereas the encoded version performs better for transversal non-Clifford gates.
Notably, we observe lower statistical distances for the preparation of the encoded magic state $\mathrm{CCZ}\ket{+++}$.
We can attribute the difference between the results for Clifford and non-Clifford gates to the compilation of the three-qubit CCZ gate into a circuit involving multiple two-qubit gates on the IonQ device~\cite{ionq-native}.
And the discrepancy between the results for $\ket{+++}$ and $\ket{\mathrm{GHZ}}$ state preparation is expected, given that the bare circuit for preparing the former requires only single-qubit gates and the latter requires two entangling gates.
We again relegate the results for $Z$ basis measurement to \cref{app:Z-basis}, as they are qualitatively similar to the results for $X$ basis measurement.

\section{Discussion}
\label{sec:discuss}

We have shown that using the \NKD{8}{3}{2} code allows us to prepare certain (encoded) states more accurately (as measured by the statistical distance) than using the native gates to prepare the same (non-encoded) states. 
We observe this advantage across a range of circuits on two different hardware platforms: IBM's superconducting qubits and IonQ's trapped-ion qubits.
The all-to-all connectivity of the IonQ device that we used enabled us to run more circuits fault-tolerantly than we could on the IBM device.
In particular, we were able to interrogate the performance of the \NKD{8}{3}{2} code for preparing magic states of the form $g \ket{+++}$, where $g \in \mathrm{CCZ} \times \{ I, \mathrm{CZ}_{12}, \mathrm{CZ}_{13}, \mathrm{CZ}_{23} \}$.
We observe an improved performance for the encoded version of circuits for preparing these states, illustrating the utility of codes like the \NKD{8}{3}{2} code, where multi-qubit non-Clifford gates can be applied using single-qubit operations.

The \NKD{8}{3}{2} is one example of a family of codes, known as generalized triorthogonal codes~\cite{Campbell2017Synth,campbellUnifiedFrameworkMagic2017,haahCodesProtocolsDistilling2018}, with transversal mult-qubit $Z$ rotations implemented by single-qubit gates.
In future it would be interesting to test the performance of larger codes in this family with higher distance.
For example, Ref.~\cite{haahCodesProtocolsDistilling2018} gives a \NKD{64}{6}{4} code with a transversal $\mathrm{CCZ}^{\otimes 2}$ gate and it is possible that smaller examples could be found using the techniques of~\cite{nezamiClassificationSmallTriorthogonal2022,Hu2022designingquantum,webster2023TransversalDiagonalLogical}.
In addition, 3D color codes~\cite{bombin2007topological,bombin2015gauge,kubica2015universal} are also generalized triorthogonal codes and therefore our approach could be extended to to the error-correcting regime by exchanging the \NKD{8}{3}{2} code for a color code with larger distance (for concrete examples see~\cite{vasmer2019three,vasmer2022morphing}).

As with any stabilizer code, the transversal gates of the \NKD{8}{3}{2} code do not form a universal set of gates.
Therefore, in order to use the \NKD{8}{3}{2} code or a similar code to implement an actual quantum algorithm, we would need to supplement the transversal gates with additional fault-tolerant gates in order to obtain a universal gate set.
One possibility worth considering would be to explore the implementation of logical gates via permutations of the physical qubits~\cite{grassl2013auto,chaoFaulttolerantQuantumComputation2018}, which can be fault-tolerant if implemented by qubit relabelling or physically moving the qubits.

\section*{Acknowledgements}

Research at Perimeter Institute is supported in part by the Government of Canada through the Department of Innovation, Science and Economic Development Canada and by the Province of Ontario through the Ministry of Colleges and Universities.
We acknowledge the support of the Natural Sciences and Engineering Research Council of Canada (NSERC).
We thank IonQ for giving us access to their hardware through the IonQ Research Credits Program.
We acknowledge CMC Microsystems for facilitating this research, specifically through their member access to the IBM Quantum Hub at PINQ$^2$.
We thank Benjamin Brown, Ben Criger, Joel Klassen and James Seddon for useful discussions.
We thank Raymond Laflamme for comments on an earlier version of this manuscript.

\bibliography{main}

\begin{thebibliography}{53}
\expandafter\ifx\csname natexlab\endcsname\relax\def\natexlab#1{#1}\fi
\expandafter\ifx\csname bibnamefont\endcsname\relax
  \def\bibnamefont#1{#1}\fi
\expandafter\ifx\csname bibfnamefont\endcsname\relax
  \def\bibfnamefont#1{#1}\fi
\expandafter\ifx\csname citenamefont\endcsname\relax
  \def\citenamefont#1{#1}\fi
\expandafter\ifx\csname url\endcsname\relax
  \def\url#1{\texttt{#1}}\fi
\expandafter\ifx\csname urlprefix\endcsname\relax\def\urlprefix{URL }\fi
\providecommand{\bibinfo}[2]{#2}
\providecommand{\eprint}[2][]{\url{#2}}

\bibitem[{\citenamefont{Shor}(1995)}]{shorSchemeReducingDecoherence1995}
\bibinfo{author}{\bibfnamefont{P.~W.} \bibnamefont{Shor}}, \bibinfo{journal}{Phys. Rev. A} \textbf{\bibinfo{volume}{52}}, \bibinfo{pages}{R2493} (\bibinfo{year}{1995}).

\bibitem[{\citenamefont{Steane}(1996{\natexlab{a}})}]{Steane1996QEC}
\bibinfo{author}{\bibfnamefont{A.~M.} \bibnamefont{Steane}}, \bibinfo{journal}{Phys. Rev. Lett.} \textbf{\bibinfo{volume}{77}}, \bibinfo{pages}{793} (\bibinfo{year}{1996}{\natexlab{a}}).

\bibitem[{\citenamefont{Chuang and Laflamme}(1995)}]{chuang1995QuantumErrorCorrection}
\bibinfo{author}{\bibfnamefont{I.~L.} \bibnamefont{Chuang}} \bibnamefont{and} \bibinfo{author}{\bibfnamefont{R.}~\bibnamefont{Laflamme}}, \bibinfo{journal}{arXiv preprint}  (\bibinfo{year}{1995}), \eprint{arXiv:quant-ph/9511003}.

\bibitem[{\citenamefont{Shor}(1996)}]{shorFaulttolerantQuantumComputation1996}
\bibinfo{author}{\bibfnamefont{P.}~\bibnamefont{Shor}}, in \emph{\bibinfo{booktitle}{Proceedings of 37th {{Conference}} on {{Foundations}} of {{Computer Science}}}} (\bibinfo{year}{1996}), pp. \bibinfo{pages}{56--65}, ISBN \bibinfo{isbn}{978-0-8186-7594-2}.

\bibitem[{\citenamefont{Eastin and Knill}(2009)}]{eastinRestrictionsTransversalEncoded2009}
\bibinfo{author}{\bibfnamefont{B.}~\bibnamefont{Eastin}} \bibnamefont{and} \bibinfo{author}{\bibfnamefont{E.}~\bibnamefont{Knill}}, \bibinfo{journal}{Phys. Rev. Lett.} \textbf{\bibinfo{volume}{102}}, \bibinfo{pages}{110502} (\bibinfo{year}{2009}).

\bibitem[{\citenamefont{Nebe et~al.}(2001)\citenamefont{Nebe, Rains, and Sloane}}]{nebeInvariantsCliffordGroups2001}
\bibinfo{author}{\bibfnamefont{G.}~\bibnamefont{Nebe}}, \bibinfo{author}{\bibfnamefont{E.~M.} \bibnamefont{Rains}}, \bibnamefont{and} \bibinfo{author}{\bibfnamefont{N.~J.~A.} \bibnamefont{Sloane}}, \bibinfo{journal}{Designs, Codes and Cryptography} \textbf{\bibinfo{volume}{24}}, \bibinfo{pages}{99} (\bibinfo{year}{2001}).

\bibitem[{\citenamefont{Bravyi and Kitaev}(2005)}]{bravyiUniversalQuantumComputation2005}
\bibinfo{author}{\bibfnamefont{S.}~\bibnamefont{Bravyi}} \bibnamefont{and} \bibinfo{author}{\bibfnamefont{A.}~\bibnamefont{Kitaev}}, \bibinfo{journal}{Phys. Rev. A} \textbf{\bibinfo{volume}{71}}, \bibinfo{pages}{022316} (\bibinfo{year}{2005}).

\bibitem[{\citenamefont{Knill}(2004)}]{knillFaultTolerantPostselectedQuantum2004a}
\bibinfo{author}{\bibfnamefont{E.}~\bibnamefont{Knill}}, \bibinfo{journal}{arXiv preprint}  (\bibinfo{year}{2004}), \eprint{arXiv:quant-ph/0402171}.

\bibitem[{\citenamefont{Bravyi and Haah}(2012)}]{bravyiMagicstateDistillationLow2012}
\bibinfo{author}{\bibfnamefont{S.}~\bibnamefont{Bravyi}} \bibnamefont{and} \bibinfo{author}{\bibfnamefont{J.}~\bibnamefont{Haah}}, \bibinfo{journal}{Phys. Rev. A} \textbf{\bibinfo{volume}{86}}, \bibinfo{pages}{052329} (\bibinfo{year}{2012}).

\bibitem[{\citenamefont{Kubica et~al.}(2015)\citenamefont{Kubica, Yoshida, and Pastawski}}]{kubicaUnfoldingColorCode2015}
\bibinfo{author}{\bibfnamefont{A.}~\bibnamefont{Kubica}}, \bibinfo{author}{\bibfnamefont{B.}~\bibnamefont{Yoshida}}, \bibnamefont{and} \bibinfo{author}{\bibfnamefont{F.}~\bibnamefont{Pastawski}}, \bibinfo{journal}{New J. Phys.} \textbf{\bibinfo{volume}{17}}, \bibinfo{pages}{083026} (\bibinfo{year}{2015}).

\bibitem[{\citenamefont{Campbell}(2016)}]{campbell2016blog}
\bibinfo{author}{\bibfnamefont{E.}~\bibnamefont{Campbell}}, \emph{\bibinfo{title}{The smallest interesting colour code}}, \bibinfo{howpublished}{\url{https://earltcampbell.com/2016/09/26/the-smallest-interesting-colour-code/}} (\bibinfo{year}{2016}).

\bibitem[{\citenamefont{Ballance et~al.}(2016)\citenamefont{Ballance, Harty, Linke, Sepiol, and Lucas}}]{Ballance2016HighFidelity}
\bibinfo{author}{\bibfnamefont{C.~J.} \bibnamefont{Ballance}}, \bibinfo{author}{\bibfnamefont{T.~P.} \bibnamefont{Harty}}, \bibinfo{author}{\bibfnamefont{N.~M.} \bibnamefont{Linke}}, \bibinfo{author}{\bibfnamefont{M.~A.} \bibnamefont{Sepiol}}, \bibnamefont{and} \bibinfo{author}{\bibfnamefont{D.~M.} \bibnamefont{Lucas}}, \bibinfo{journal}{Phys. Rev. Lett.} \textbf{\bibinfo{volume}{117}}, \bibinfo{pages}{060504} (\bibinfo{year}{2016}).

\bibitem[{\citenamefont{Wright et~al.}(2019)\citenamefont{Wright, Beck, Debnath, Amini, Nam, Grzesiak, Chen, Pisenti, Chmielewski, Collins et~al.}}]{wright2019Benchmarking11qubitQuantum}
\bibinfo{author}{\bibfnamefont{K.}~\bibnamefont{Wright}}, \bibinfo{author}{\bibfnamefont{K.~M.} \bibnamefont{Beck}}, \bibinfo{author}{\bibfnamefont{S.}~\bibnamefont{Debnath}}, \bibinfo{author}{\bibfnamefont{J.~M.} \bibnamefont{Amini}}, \bibinfo{author}{\bibfnamefont{Y.}~\bibnamefont{Nam}}, \bibinfo{author}{\bibfnamefont{N.}~\bibnamefont{Grzesiak}}, \bibinfo{author}{\bibfnamefont{J.-S.} \bibnamefont{Chen}}, \bibinfo{author}{\bibfnamefont{N.~C.} \bibnamefont{Pisenti}}, \bibinfo{author}{\bibfnamefont{M.}~\bibnamefont{Chmielewski}}, \bibinfo{author}{\bibfnamefont{C.}~\bibnamefont{Collins}}, \bibnamefont{et~al.}, \bibinfo{journal}{Nat. Commun.} \textbf{\bibinfo{volume}{10}}, \bibinfo{pages}{5464} (\bibinfo{year}{2019}).

\bibitem[{\citenamefont{Jurcevic et~al.}(2021)\citenamefont{Jurcevic, {Javadi-Abhari}, Bishop, Lauer, Bogorin, Brink, Capelluto, G{\"u}nl{\"u}k, Itoko, Kanazawa et~al.}}]{jurcevic2021DemonstrationQuantumVolume}
\bibinfo{author}{\bibfnamefont{P.}~\bibnamefont{Jurcevic}}, \bibinfo{author}{\bibfnamefont{A.}~\bibnamefont{{Javadi-Abhari}}}, \bibinfo{author}{\bibfnamefont{L.~S.} \bibnamefont{Bishop}}, \bibinfo{author}{\bibfnamefont{I.}~\bibnamefont{Lauer}}, \bibinfo{author}{\bibfnamefont{D.~F.} \bibnamefont{Bogorin}}, \bibinfo{author}{\bibfnamefont{M.}~\bibnamefont{Brink}}, \bibinfo{author}{\bibfnamefont{L.}~\bibnamefont{Capelluto}}, \bibinfo{author}{\bibfnamefont{O.}~\bibnamefont{G{\"u}nl{\"u}k}}, \bibinfo{author}{\bibfnamefont{T.}~\bibnamefont{Itoko}}, \bibinfo{author}{\bibfnamefont{N.}~\bibnamefont{Kanazawa}}, \bibnamefont{et~al.}, \bibinfo{journal}{Quantum Sci. Technol.} \textbf{\bibinfo{volume}{6}}, \bibinfo{pages}{025020} (\bibinfo{year}{2021}).

\bibitem[{\citenamefont{Wu et~al.}(2021)\citenamefont{Wu, Bao, Cao, Chen, Chen, Chen, Chung, Deng, Du, Fan et~al.}}]{Wu2021QuantumAdvantage}
\bibinfo{author}{\bibfnamefont{Y.}~\bibnamefont{Wu}}, \bibinfo{author}{\bibfnamefont{W.-S.} \bibnamefont{Bao}}, \bibinfo{author}{\bibfnamefont{S.}~\bibnamefont{Cao}}, \bibinfo{author}{\bibfnamefont{F.}~\bibnamefont{Chen}}, \bibinfo{author}{\bibfnamefont{M.-C.} \bibnamefont{Chen}}, \bibinfo{author}{\bibfnamefont{X.}~\bibnamefont{Chen}}, \bibinfo{author}{\bibfnamefont{T.-H.} \bibnamefont{Chung}}, \bibinfo{author}{\bibfnamefont{H.}~\bibnamefont{Deng}}, \bibinfo{author}{\bibfnamefont{Y.}~\bibnamefont{Du}}, \bibinfo{author}{\bibfnamefont{D.}~\bibnamefont{Fan}}, \bibnamefont{et~al.}, \bibinfo{journal}{Phys. Rev. Lett.} \textbf{\bibinfo{volume}{127}}, \bibinfo{pages}{180501} (\bibinfo{year}{2021}).

\bibitem[{\citenamefont{Huang et~al.}(2019)\citenamefont{Huang, Yang, Chan, Tanttu, Hensen, Leon, Fogarty, Hwang, Hudson, Itoh et~al.}}]{huang2019FidelityBenchmarksTwoqubit}
\bibinfo{author}{\bibfnamefont{W.}~\bibnamefont{Huang}}, \bibinfo{author}{\bibfnamefont{C.~H.} \bibnamefont{Yang}}, \bibinfo{author}{\bibfnamefont{K.~W.} \bibnamefont{Chan}}, \bibinfo{author}{\bibfnamefont{T.}~\bibnamefont{Tanttu}}, \bibinfo{author}{\bibfnamefont{B.}~\bibnamefont{Hensen}}, \bibinfo{author}{\bibfnamefont{R.~C.~C.} \bibnamefont{Leon}}, \bibinfo{author}{\bibfnamefont{M.~A.} \bibnamefont{Fogarty}}, \bibinfo{author}{\bibfnamefont{J.~C.~C.} \bibnamefont{Hwang}}, \bibinfo{author}{\bibfnamefont{F.~E.} \bibnamefont{Hudson}}, \bibinfo{author}{\bibfnamefont{K.~M.} \bibnamefont{Itoh}}, \bibnamefont{et~al.}, \bibinfo{journal}{Nature} \textbf{\bibinfo{volume}{569}}, \bibinfo{pages}{532} (\bibinfo{year}{2019}).

\bibitem[{\citenamefont{Moses et~al.}(2023)\citenamefont{Moses, Baldwin, Allman, Ancona, Ascarrunz, Barnes, Bartolotta, Bjork, Blanchard, Bohn et~al.}}]{moses2023RaceTrackTrappedIon}
\bibinfo{author}{\bibfnamefont{S.~A.} \bibnamefont{Moses}}, \bibinfo{author}{\bibfnamefont{C.~H.} \bibnamefont{Baldwin}}, \bibinfo{author}{\bibfnamefont{M.~S.} \bibnamefont{Allman}}, \bibinfo{author}{\bibfnamefont{R.}~\bibnamefont{Ancona}}, \bibinfo{author}{\bibfnamefont{L.}~\bibnamefont{Ascarrunz}}, \bibinfo{author}{\bibfnamefont{C.}~\bibnamefont{Barnes}}, \bibinfo{author}{\bibfnamefont{J.}~\bibnamefont{Bartolotta}}, \bibinfo{author}{\bibfnamefont{B.}~\bibnamefont{Bjork}}, \bibinfo{author}{\bibfnamefont{P.}~\bibnamefont{Blanchard}}, \bibinfo{author}{\bibfnamefont{M.}~\bibnamefont{Bohn}}, \bibnamefont{et~al.}, \bibinfo{journal}{arXiv preprint}  (\bibinfo{year}{2023}), \eprint{arXiv:2305.03828}.

\bibitem[{\citenamefont{Gottesman}(2016)}]{gottesman_quantum_2016}
\bibinfo{author}{\bibfnamefont{D.}~\bibnamefont{Gottesman}}, \bibinfo{journal}{arXiv preprint}  (\bibinfo{year}{2016}), \eprint{arXiv:1610.03507}.

\bibitem[{\citenamefont{Souza et~al.}(2011)\citenamefont{Souza, Zhang, Ryan, and Laflamme}}]{souza2011ExperimentalMagicState}
\bibinfo{author}{\bibfnamefont{A.~M.} \bibnamefont{Souza}}, \bibinfo{author}{\bibfnamefont{J.}~\bibnamefont{Zhang}}, \bibinfo{author}{\bibfnamefont{C.~A.} \bibnamefont{Ryan}}, \bibnamefont{and} \bibinfo{author}{\bibfnamefont{R.}~\bibnamefont{Laflamme}}, \bibinfo{journal}{Nat. Commun.} \textbf{\bibinfo{volume}{2}}, \bibinfo{pages}{169} (\bibinfo{year}{2011}).

\bibitem[{\citenamefont{Zhang et~al.}(2012)\citenamefont{Zhang, Laflamme, and Suter}}]{Zhang2012Experimental}
\bibinfo{author}{\bibfnamefont{J.}~\bibnamefont{Zhang}}, \bibinfo{author}{\bibfnamefont{R.}~\bibnamefont{Laflamme}}, \bibnamefont{and} \bibinfo{author}{\bibfnamefont{D.}~\bibnamefont{Suter}}, \bibinfo{journal}{Phys. Rev. Lett.} \textbf{\bibinfo{volume}{109}}, \bibinfo{pages}{100503} (\bibinfo{year}{2012}).

\bibitem[{\citenamefont{Nigg et~al.}(2014)\citenamefont{Nigg, Mueller, Martinez, Schindler, Hennrich, Monz, {Martin-Delgado}, and Blatt}}]{niggExperimentalQuantumComputations2014}
\bibinfo{author}{\bibfnamefont{D.}~\bibnamefont{Nigg}}, \bibinfo{author}{\bibfnamefont{M.}~\bibnamefont{Mueller}}, \bibinfo{author}{\bibfnamefont{E.~A.} \bibnamefont{Martinez}}, \bibinfo{author}{\bibfnamefont{P.}~\bibnamefont{Schindler}}, \bibinfo{author}{\bibfnamefont{M.}~\bibnamefont{Hennrich}}, \bibinfo{author}{\bibfnamefont{T.}~\bibnamefont{Monz}}, \bibinfo{author}{\bibfnamefont{M.~A.} \bibnamefont{{Martin-Delgado}}}, \bibnamefont{and} \bibinfo{author}{\bibfnamefont{R.}~\bibnamefont{Blatt}}, \bibinfo{journal}{Science} \textbf{\bibinfo{volume}{345}}, \bibinfo{pages}{302} (\bibinfo{year}{2014}), \eprint{arXiv:1403.5426}.

\bibitem[{\citenamefont{Egan et~al.}(2021)\citenamefont{Egan, Debroy, Noel, Risinger, Zhu, Biswas, Newman, Li, Brown, Cetina et~al.}}]{eganFaultTolerantOperationQuantum2021}
\bibinfo{author}{\bibfnamefont{L.}~\bibnamefont{Egan}}, \bibinfo{author}{\bibfnamefont{D.~M.} \bibnamefont{Debroy}}, \bibinfo{author}{\bibfnamefont{C.}~\bibnamefont{Noel}}, \bibinfo{author}{\bibfnamefont{A.}~\bibnamefont{Risinger}}, \bibinfo{author}{\bibfnamefont{D.}~\bibnamefont{Zhu}}, \bibinfo{author}{\bibfnamefont{D.}~\bibnamefont{Biswas}}, \bibinfo{author}{\bibfnamefont{M.}~\bibnamefont{Newman}}, \bibinfo{author}{\bibfnamefont{M.}~\bibnamefont{Li}}, \bibinfo{author}{\bibfnamefont{K.~R.} \bibnamefont{Brown}}, \bibinfo{author}{\bibfnamefont{M.}~\bibnamefont{Cetina}}, \bibnamefont{et~al.}, \bibinfo{journal}{arXiv preprint}  (\bibinfo{year}{2021}), \eprint{arXiv:2009.11482}.

\bibitem[{\citenamefont{Postler et~al.}(2022)\citenamefont{Postler, Heu{$\beta$}en, Pogorelov, Rispler, Feldker, Meth, Marciniak, Stricker, Ringbauer, Blatt et~al.}}]{postler2022DemonstrationFaulttolerantUniversal}
\bibinfo{author}{\bibfnamefont{L.}~\bibnamefont{Postler}}, \bibinfo{author}{\bibfnamefont{S.}~\bibnamefont{Heu{$\beta$}en}}, \bibinfo{author}{\bibfnamefont{I.}~\bibnamefont{Pogorelov}}, \bibinfo{author}{\bibfnamefont{M.}~\bibnamefont{Rispler}}, \bibinfo{author}{\bibfnamefont{T.}~\bibnamefont{Feldker}}, \bibinfo{author}{\bibfnamefont{M.}~\bibnamefont{Meth}}, \bibinfo{author}{\bibfnamefont{C.~D.} \bibnamefont{Marciniak}}, \bibinfo{author}{\bibfnamefont{R.}~\bibnamefont{Stricker}}, \bibinfo{author}{\bibfnamefont{M.}~\bibnamefont{Ringbauer}}, \bibinfo{author}{\bibfnamefont{R.}~\bibnamefont{Blatt}}, \bibnamefont{et~al.}, \bibinfo{journal}{Nature} \textbf{\bibinfo{volume}{605}}, \bibinfo{pages}{675} (\bibinfo{year}{2022}).

\bibitem[{\citenamefont{{Ryan-Anderson} et~al.}(2022)\citenamefont{{Ryan-Anderson}, Brown, Allman, Arkin, {Asa-Attuah}, Baldwin, Berg, Bohnet, Braxton, Burdick et~al.}}]{ryan-anderson2022ImplementingFaulttolerantEntangling}
\bibinfo{author}{\bibfnamefont{C.}~\bibnamefont{{Ryan-Anderson}}}, \bibinfo{author}{\bibfnamefont{N.~C.} \bibnamefont{Brown}}, \bibinfo{author}{\bibfnamefont{M.~S.} \bibnamefont{Allman}}, \bibinfo{author}{\bibfnamefont{B.}~\bibnamefont{Arkin}}, \bibinfo{author}{\bibfnamefont{G.}~\bibnamefont{{Asa-Attuah}}}, \bibinfo{author}{\bibfnamefont{C.}~\bibnamefont{Baldwin}}, \bibinfo{author}{\bibfnamefont{J.}~\bibnamefont{Berg}}, \bibinfo{author}{\bibfnamefont{J.~G.} \bibnamefont{Bohnet}}, \bibinfo{author}{\bibfnamefont{S.}~\bibnamefont{Braxton}}, \bibinfo{author}{\bibfnamefont{N.}~\bibnamefont{Burdick}}, \bibnamefont{et~al.}, \bibinfo{journal}{arXiv preprint}  (\bibinfo{year}{2022}), \eprint{arXiv:2208.01863}.

\bibitem[{\citenamefont{Vuillot}(2018)}]{vuillotErrorDetectionHelpful2018}
\bibinfo{author}{\bibfnamefont{C.}~\bibnamefont{Vuillot}}, \bibinfo{journal}{Quantum Inf. Comput.} \textbf{\bibinfo{volume}{18}}, \bibinfo{pages}{0949} (\bibinfo{year}{2018}).

\bibitem[{\citenamefont{Harper and Flammia}(2019)}]{harperFaultTolerantLogicalGates2019}
\bibinfo{author}{\bibfnamefont{R.}~\bibnamefont{Harper}} \bibnamefont{and} \bibinfo{author}{\bibfnamefont{S.~T.} \bibnamefont{Flammia}}, \bibinfo{journal}{Phys. Rev. Lett.} \textbf{\bibinfo{volume}{122}}, \bibinfo{pages}{080504} (\bibinfo{year}{2019}).

\bibitem[{\citenamefont{Gupta et~al.}(2023)\citenamefont{Gupta, Sundaresan, Alexander, Wood, Merkel, Healy, Hillenbrand, {Jochym-O'Connor}, Wootton, Yoder et~al.}}]{gupta2023EncodingMagicState}
\bibinfo{author}{\bibfnamefont{R.~S.} \bibnamefont{Gupta}}, \bibinfo{author}{\bibfnamefont{N.}~\bibnamefont{Sundaresan}}, \bibinfo{author}{\bibfnamefont{T.}~\bibnamefont{Alexander}}, \bibinfo{author}{\bibfnamefont{C.~J.} \bibnamefont{Wood}}, \bibinfo{author}{\bibfnamefont{S.~T.} \bibnamefont{Merkel}}, \bibinfo{author}{\bibfnamefont{M.~B.} \bibnamefont{Healy}}, \bibinfo{author}{\bibfnamefont{M.}~\bibnamefont{Hillenbrand}}, \bibinfo{author}{\bibfnamefont{T.}~\bibnamefont{{Jochym-O'Connor}}}, \bibinfo{author}{\bibfnamefont{J.~R.} \bibnamefont{Wootton}}, \bibinfo{author}{\bibfnamefont{T.~J.} \bibnamefont{Yoder}}, \bibnamefont{et~al.}, \bibinfo{journal}{arXiv preprint}  (\bibinfo{year}{2023}), \eprint{arXiv:2305.13581}.

\bibitem[{\citenamefont{Vedral et~al.}(1996)\citenamefont{Vedral, Barenco, and Ekert}}]{Vedral1996QArithmetic}
\bibinfo{author}{\bibfnamefont{V.}~\bibnamefont{Vedral}}, \bibinfo{author}{\bibfnamefont{A.}~\bibnamefont{Barenco}}, \bibnamefont{and} \bibinfo{author}{\bibfnamefont{A.}~\bibnamefont{Ekert}}, \bibinfo{journal}{Phys. Rev. A} \textbf{\bibinfo{volume}{54}}, \bibinfo{pages}{147} (\bibinfo{year}{1996}).

\bibitem[{\citenamefont{Gidney}(2018)}]{Gidney2018halvingcostof}
\bibinfo{author}{\bibfnamefont{C.}~\bibnamefont{Gidney}}, \bibinfo{journal}{{Quantum}} \textbf{\bibinfo{volume}{2}}, \bibinfo{pages}{74} (\bibinfo{year}{2018}).

\bibitem[{\citenamefont{Shor}(1997)}]{shorPolynomialTimeAlgorithmsPrime1997}
\bibinfo{author}{\bibfnamefont{P.~W.} \bibnamefont{Shor}}, \bibinfo{journal}{SIAM J. Comput.} \textbf{\bibinfo{volume}{26}}, \bibinfo{pages}{1484} (\bibinfo{year}{1997}).

\bibitem[{\citenamefont{Gidney and Eker{\aa{}}}(2021)}]{Gidney2021howtofactorbit}
\bibinfo{author}{\bibfnamefont{C.}~\bibnamefont{Gidney}} \bibnamefont{and} \bibinfo{author}{\bibfnamefont{M.}~\bibnamefont{Eker{\aa{}}}}, \bibinfo{journal}{{Quantum}} \textbf{\bibinfo{volume}{5}}, \bibinfo{pages}{433} (\bibinfo{year}{2021}).

\bibitem[{\citenamefont{Gottesman}(1997)}]{gottesmanStabilizerCodesQuantum1997}
\bibinfo{author}{\bibfnamefont{D.}~\bibnamefont{Gottesman}}, Ph.D. thesis, \bibinfo{school}{Caltech} (\bibinfo{year}{1997}), \eprint{arXiv:quant-ph/9705052}.

\bibitem[{\citenamefont{Calderbank and Shor}(1996)}]{calderbankGoodQuantumErrorcorrecting1996}
\bibinfo{author}{\bibfnamefont{A.~R.} \bibnamefont{Calderbank}} \bibnamefont{and} \bibinfo{author}{\bibfnamefont{P.~W.} \bibnamefont{Shor}}, \bibinfo{journal}{Phys. Rev. A} \textbf{\bibinfo{volume}{54}}, \bibinfo{pages}{1098} (\bibinfo{year}{1996}).

\bibitem[{\citenamefont{Steane}(1996{\natexlab{b}})}]{steaneMultipleparticleInterferenceQuantum1996}
\bibinfo{author}{\bibfnamefont{A.~M.} \bibnamefont{Steane}}, \bibinfo{journal}{Proc. R. Soc. Lond. A} \textbf{\bibinfo{volume}{452}}, \bibinfo{pages}{2551} (\bibinfo{year}{1996}{\natexlab{b}}).

\bibitem[{\citenamefont{Nielsen and Chuang}(2010)}]{nielsenchuang}
\bibinfo{author}{\bibfnamefont{M.~A.} \bibnamefont{Nielsen}} \bibnamefont{and} \bibinfo{author}{\bibfnamefont{I.~L.} \bibnamefont{Chuang}}, \emph{\bibinfo{title}{Quantum Computation and Quantum Information: 10th Anniversary Edition}} (\bibinfo{publisher}{Cambridge University Press}, \bibinfo{year}{2010}).

\bibitem[{\citenamefont{Chao and Reichardt}(2018{\natexlab{a}})}]{chao_quantum_2018}
\bibinfo{author}{\bibfnamefont{R.}~\bibnamefont{Chao}} \bibnamefont{and} \bibinfo{author}{\bibfnamefont{B.~W.} \bibnamefont{Reichardt}}, \bibinfo{journal}{Phys. Rev. Lett.} \textbf{\bibinfo{volume}{121}}, \bibinfo{pages}{050502} (\bibinfo{year}{2018}{\natexlab{a}}).

\bibitem[{ibm()}]{ibmq}
\emph{\bibinfo{title}{{IBM} {Q}uantum}}, \bibinfo{howpublished}{\url{https://quantum-computing.ibm.com/}}.

\bibitem[{\citenamefont{Hertzberg et~al.}(2021)\citenamefont{Hertzberg, Zhang, Rosenblatt, Magesan, Smolin, Yau, Adiga, Sandberg, Brink, Chow et~al.}}]{hertzberg2021LaserannealingJosephsonJunctions}
\bibinfo{author}{\bibfnamefont{J.~B.} \bibnamefont{Hertzberg}}, \bibinfo{author}{\bibfnamefont{E.~J.} \bibnamefont{Zhang}}, \bibinfo{author}{\bibfnamefont{S.}~\bibnamefont{Rosenblatt}}, \bibinfo{author}{\bibfnamefont{E.}~\bibnamefont{Magesan}}, \bibinfo{author}{\bibfnamefont{J.~A.} \bibnamefont{Smolin}}, \bibinfo{author}{\bibfnamefont{J.-B.} \bibnamefont{Yau}}, \bibinfo{author}{\bibfnamefont{V.~P.} \bibnamefont{Adiga}}, \bibinfo{author}{\bibfnamefont{M.}~\bibnamefont{Sandberg}}, \bibinfo{author}{\bibfnamefont{M.}~\bibnamefont{Brink}}, \bibinfo{author}{\bibfnamefont{J.~M.} \bibnamefont{Chow}}, \bibnamefont{et~al.}, \bibinfo{journal}{npj Quantum Inf.} \textbf{\bibinfo{volume}{7}}, \bibinfo{pages}{129} (\bibinfo{year}{2021}).

\bibitem[{\citenamefont{Vasmer}()}]{github}
\bibinfo{author}{\bibfnamefont{M.}~\bibnamefont{Vasmer}}, \emph{\bibinfo{title}{Data analysis for "implementing fault-tolerant non-clifford gates using the [[8,3,2]] color code"}}, \bibinfo{howpublished}{\url{https://github.com/MikeVasmer/832-code}}.

\bibitem[{ion()}]{ionq-native}
\emph{\bibinfo{title}{Getting started with {N}ative {G}ates}}, \bibinfo{howpublished}{\url{https://ionq.com/docs/getting-started-with-native-gates}}.

\bibitem[{\citenamefont{Campbell and Howard}(2017{\natexlab{a}})}]{Campbell2017Synth}
\bibinfo{author}{\bibfnamefont{E.~T.} \bibnamefont{Campbell}} \bibnamefont{and} \bibinfo{author}{\bibfnamefont{M.}~\bibnamefont{Howard}}, \bibinfo{journal}{Phys. Rev. Lett.} \textbf{\bibinfo{volume}{118}}, \bibinfo{pages}{060501} (\bibinfo{year}{2017}{\natexlab{a}}).

\bibitem[{\citenamefont{Campbell and Howard}(2017{\natexlab{b}})}]{campbellUnifiedFrameworkMagic2017}
\bibinfo{author}{\bibfnamefont{E.~T.} \bibnamefont{Campbell}} \bibnamefont{and} \bibinfo{author}{\bibfnamefont{M.}~\bibnamefont{Howard}}, \bibinfo{journal}{Phys. Rev. A} \textbf{\bibinfo{volume}{95}}, \bibinfo{pages}{022316} (\bibinfo{year}{2017}{\natexlab{b}}).

\bibitem[{\citenamefont{Haah and Hastings}(2018)}]{haahCodesProtocolsDistilling2018}
\bibinfo{author}{\bibfnamefont{J.}~\bibnamefont{Haah}} \bibnamefont{and} \bibinfo{author}{\bibfnamefont{M.~B.} \bibnamefont{Hastings}}, \bibinfo{journal}{Quantum} \textbf{\bibinfo{volume}{2}}, \bibinfo{pages}{71} (\bibinfo{year}{2018}).

\bibitem[{\citenamefont{Nezami and Haah}(2022)}]{nezamiClassificationSmallTriorthogonal2022}
\bibinfo{author}{\bibfnamefont{S.}~\bibnamefont{Nezami}} \bibnamefont{and} \bibinfo{author}{\bibfnamefont{J.}~\bibnamefont{Haah}}, \bibinfo{journal}{Phys. Rev. A} \textbf{\bibinfo{volume}{106}}, \bibinfo{pages}{012437} (\bibinfo{year}{2022}).

\bibitem[{\citenamefont{Hu et~al.}(2022)\citenamefont{Hu, Liang, and Calderbank}}]{Hu2022designingquantum}
\bibinfo{author}{\bibfnamefont{J.}~\bibnamefont{Hu}}, \bibinfo{author}{\bibfnamefont{Q.}~\bibnamefont{Liang}}, \bibnamefont{and} \bibinfo{author}{\bibfnamefont{R.}~\bibnamefont{Calderbank}}, \bibinfo{journal}{{Quantum}} \textbf{\bibinfo{volume}{6}}, \bibinfo{pages}{802} (\bibinfo{year}{2022}).

\bibitem[{\citenamefont{Webster et~al.}(2023)\citenamefont{Webster, Quintavalle, and Bartlett}}]{webster2023TransversalDiagonalLogical}
\bibinfo{author}{\bibfnamefont{M.~A.} \bibnamefont{Webster}}, \bibinfo{author}{\bibfnamefont{A.~O.} \bibnamefont{Quintavalle}}, \bibnamefont{and} \bibinfo{author}{\bibfnamefont{S.~D.} \bibnamefont{Bartlett}}, \bibinfo{journal}{arXiv preprint}  (\bibinfo{year}{2023}), \eprint{arXiv:2303.15615}.

\bibitem[{\citenamefont{Bombin and Martin-Delgado}(2007)}]{bombin2007topological}
\bibinfo{author}{\bibfnamefont{H.}~\bibnamefont{Bombin}} \bibnamefont{and} \bibinfo{author}{\bibfnamefont{M.~A.} \bibnamefont{Martin-Delgado}}, \bibinfo{journal}{Phys. Rev. Lett.} \textbf{\bibinfo{volume}{98}}, \bibinfo{pages}{160502} (\bibinfo{year}{2007}).

\bibitem[{\citenamefont{Bombín}(2015)}]{bombin2015gauge}
\bibinfo{author}{\bibfnamefont{H.}~\bibnamefont{Bombín}}, \bibinfo{journal}{New Journal of Physics} \textbf{\bibinfo{volume}{17}}, \bibinfo{pages}{083002} (\bibinfo{year}{2015}).

\bibitem[{\citenamefont{Kubica and Beverland}(2015)}]{kubica2015universal}
\bibinfo{author}{\bibfnamefont{A.}~\bibnamefont{Kubica}} \bibnamefont{and} \bibinfo{author}{\bibfnamefont{M.~E.} \bibnamefont{Beverland}}, \bibinfo{journal}{Phys. Rev. A} \textbf{\bibinfo{volume}{91}}, \bibinfo{pages}{032330} (\bibinfo{year}{2015}).

\bibitem[{\citenamefont{Vasmer and Browne}(2019)}]{vasmer2019three}
\bibinfo{author}{\bibfnamefont{M.}~\bibnamefont{Vasmer}} \bibnamefont{and} \bibinfo{author}{\bibfnamefont{D.~E.} \bibnamefont{Browne}}, \bibinfo{journal}{Phys. Rev. A} \textbf{\bibinfo{volume}{100}}, \bibinfo{pages}{012312} (\bibinfo{year}{2019}).

\bibitem[{\citenamefont{Vasmer and Kubica}(2022)}]{vasmer2022morphing}
\bibinfo{author}{\bibfnamefont{M.}~\bibnamefont{Vasmer}} \bibnamefont{and} \bibinfo{author}{\bibfnamefont{A.}~\bibnamefont{Kubica}}, \bibinfo{journal}{PRX Quantum} \textbf{\bibinfo{volume}{3}}, \bibinfo{pages}{030319} (\bibinfo{year}{2022}).

\bibitem[{\citenamefont{Grassl and Roetteler}(2013)}]{grassl2013auto}
\bibinfo{author}{\bibfnamefont{M.}~\bibnamefont{Grassl}} \bibnamefont{and} \bibinfo{author}{\bibfnamefont{M.}~\bibnamefont{Roetteler}}, in \emph{\bibinfo{booktitle}{2013 IEEE International Symposium on Information Theory}} (\bibinfo{year}{2013}), pp. \bibinfo{pages}{534--538}.

\bibitem[{\citenamefont{Chao and Reichardt}(2018{\natexlab{b}})}]{chaoFaulttolerantQuantumComputation2018}
\bibinfo{author}{\bibfnamefont{R.}~\bibnamefont{Chao}} \bibnamefont{and} \bibinfo{author}{\bibfnamefont{B.~W.} \bibnamefont{Reichardt}}, \bibinfo{journal}{npj Quantum Inf.} \textbf{\bibinfo{volume}{4}}, \bibinfo{pages}{42} (\bibinfo{year}{2018}{\natexlab{b}}).

\end{thebibliography}

\appendix
\onecolumngrid

\section{Additional demonstration results}

\subsection{Logical \texorpdfstring{$Z$}{\emph{Z}} basis measurement results}
\label{app:Z-basis}

\begin{figure*}[ht]
\centering
    \includegraphics[width=.9\linewidth]{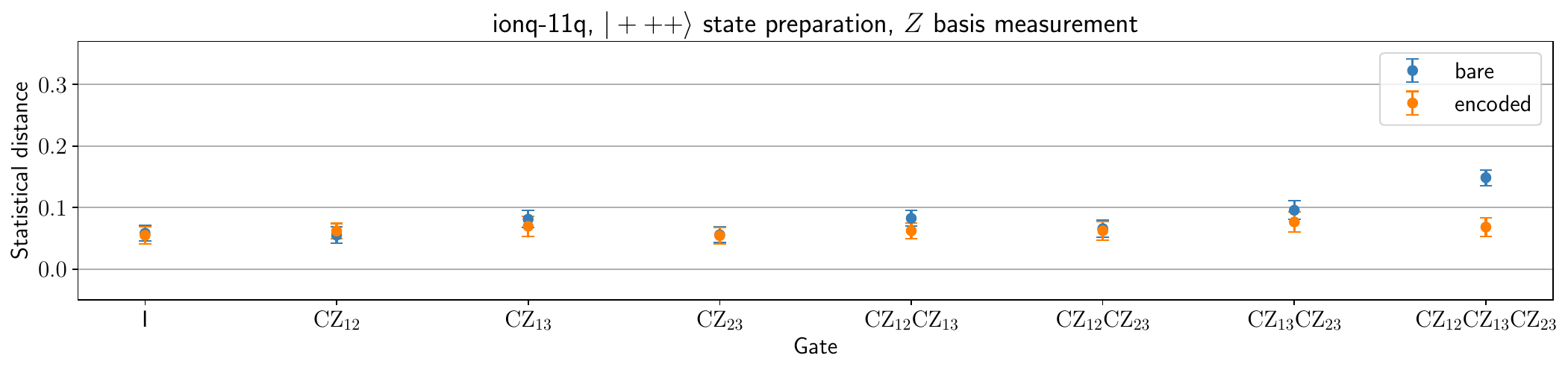}
    \includegraphics[width=.9\linewidth]{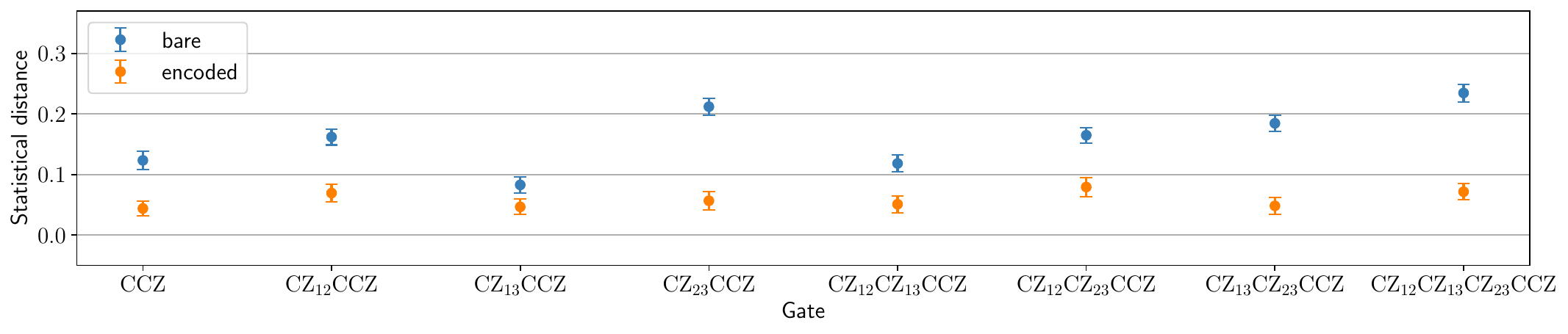}
    \caption{
    Performance of bare (non-encoded) and encoded versions of circuits for preparing states of the form $g \ket{+++}$, where $g$ is a transversal gate of the \NKD{8}{3}{2} code.
    In each case, we measure the qubits in the $Z$ basis and we plot the statistical distance of the observed measurement distribution from the ideal distribution.
    Each data point represents 1024 shots of the circuit performed on \textsf{ionq-11q}, and we use bootstrap resampling to calculate the error bars.
    }
    \label{fig:Z-+++}
\end{figure*}

\begin{figure*}[ht]
\centering
    \includegraphics[width=.9\linewidth]{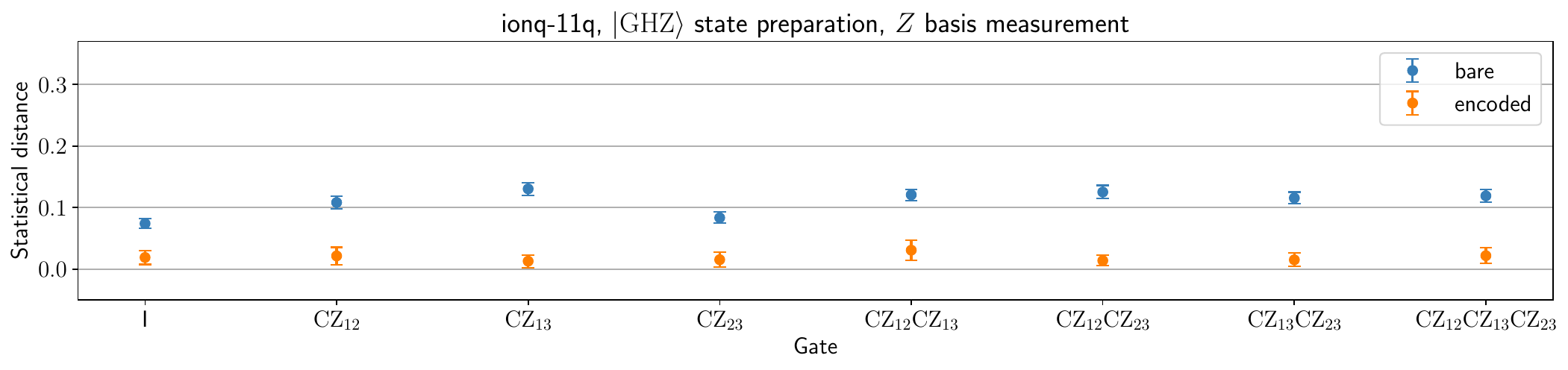}
    \includegraphics[width=.9\linewidth]{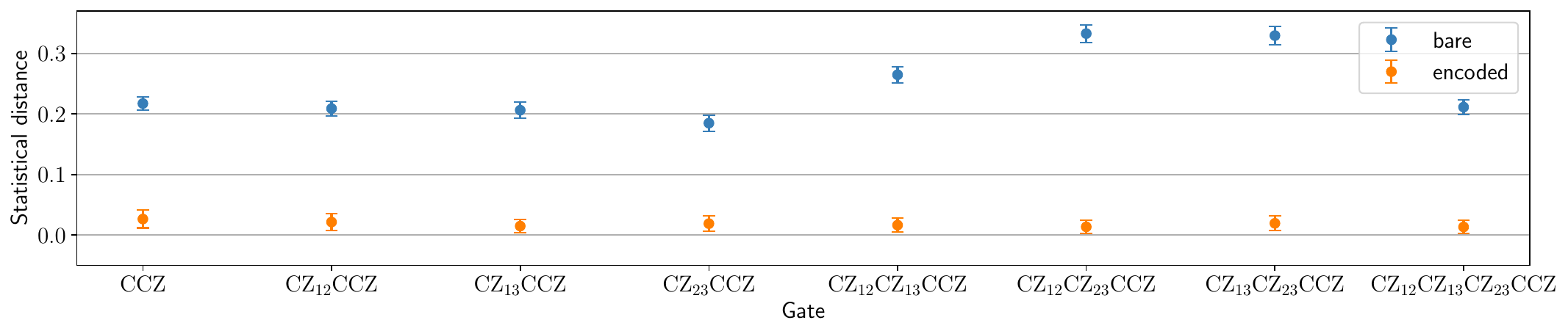}
    \includegraphics[width=.9\linewidth]{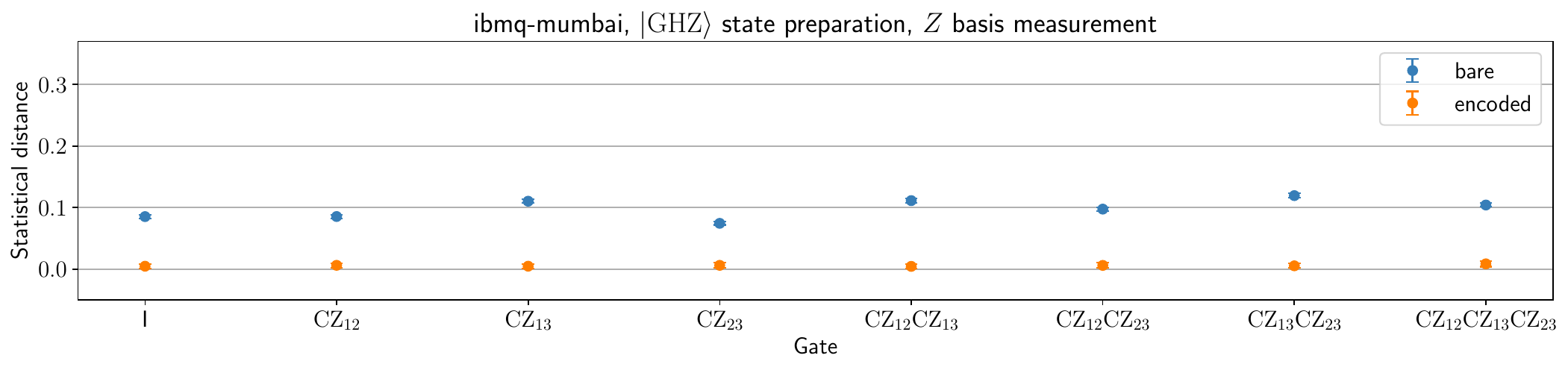}
    \includegraphics[width=.9\linewidth]{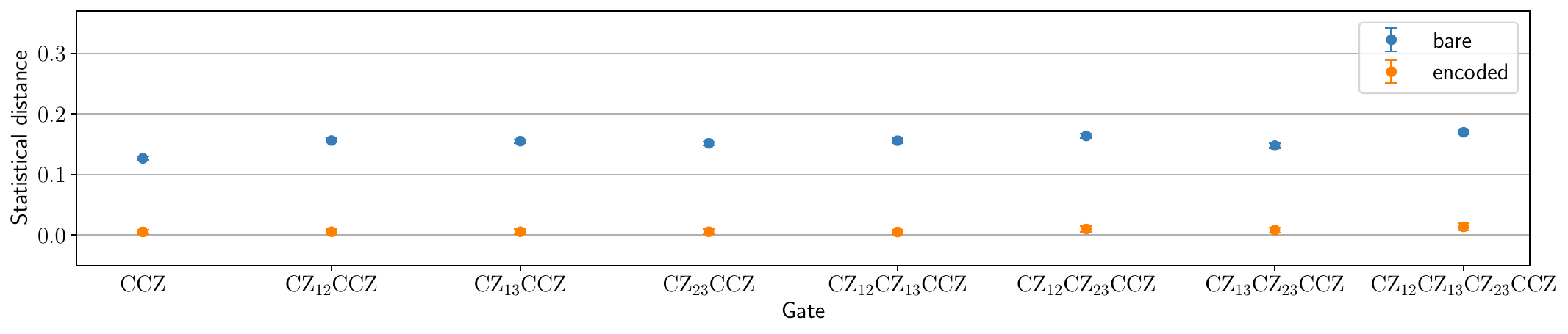}
    \caption{
    Performance of bare (non-encoded) and encoded versions of circuits for preparing states of the form $g \ket{\mathrm{GHZ}}$, where $g$ is a transversal gate of the \NKD{8}{3}{2} code.
    In each case, we measure the qubits in the $Z$ basis and we plot the statistical distance of the observed measurement distribution from the ideal distribution.
    The upper two plots show the data for \textsf{ionq-11q}, where we ran 1024 shots for each circuit, and the lower two plots show the data for \textsf{ibmq\_mumbai} where we ran 10,000 shots for each circuit.
    In both cases, the error bars are calculated using bootstrap resampling.
    }
    \label{fig:Z-GHZ}
\end{figure*}

For the demonstrations with $\ket{+++}$ state preparation and $Z$ basis measurement, we observe improved performance for the encoded circuits containing a transversal non-Clifford gate; see \cref{fig:Z-+++}.
And for the demonstrations with $\ket{\mathrm{GHZ}}$ state preparation and $Z$ basis measurement, we observe improved performance for all of the encoded circuits; see \cref{fig:Z-GHZ}.
We note that in these cases the gates have no effect on the expected output measurement distribution, as the gates commute with the measurements.

\subsection{Post-selection rates}
\label{app:post}

In this appendix, we provide the post-selection rates (proportion of accepted shots) for our demonstrations.
Two-qubit gates and measurements are the noisiest operations in superconducting and trapped-ion devices~\cite{jurcevic2021DemonstrationQuantumVolume,wright2019Benchmarking11qubitQuantum}, hence we can approximate the error model in these devices with ideal state preparation and one-qubit gates, but noisy two-qubit gates and measurements (1\% error rate).
Suppose that any two-qubit gate error or measurement error causes us to discard the run, then (to first order) we would expect a post-selection rate of $1 - 0.01 (n_m + n_g)$, where $n_m$ is the number of measurements and $n_g$ is the number of two-qubit gates.
By counting the relevant locations in \cref{fig:state-prep-ghz,fig:state-prep-+++}, we therefore estimate a post-selection rate of $69\%$ and $86\%$ for the circuits with $\ket{+++}$ and $\ket{\mathrm{GHZ}}$ state preparation, respectively.  
These estimates are reasonably close to the values we observe in our demonstrations; see \cref{tab:ionq-ghz,tab:ionq-+++,tab:ibm-+++}.

\begin{table}[ht]
\begin{center}
\begin{tabular}{ c c c c c }
\hline\hline 
State & Gate & Measurement & Device & Post-selection rate \\
\hline
$\ket{+++}$ & $I$ & $X$ basis & \textsf{ionq-11q} & 78\% \\
$\ket{+++}$ & $CZ_{12}$ & $X$ basis & \textsf{ionq-11q} & 78\% \\
$\ket{+++}$ & $CZ_{13}$ & $X$ basis & \textsf{ionq-11q} & 69\% \\
$\ket{+++}$ & $CZ_{12}CZ_{13}$ & $X$ basis & \textsf{ionq-11q} & 76\% \\
$\ket{+++}$ & $CZ_{23}$ & $X$ basis & \textsf{ionq-11q} & 80\% \\
$\ket{+++}$ & $CZ_{12}CZ_{23}$ & $X$ basis & \textsf{ionq-11q} & 77\% \\
$\ket{+++}$ & $CZ_{13}CZ_{23}$ & $X$ basis & \textsf{ionq-11q} & 79\% \\
$\ket{+++}$ & $CZ_{12}CZ_{13}CZ_{23}$ & $X$ basis & \textsf{ionq-11q} & 80\% \\
$\ket{+++}$ & $CCZ$ & $X$ basis & \textsf{ionq-11q} & 75\% \\
$\ket{+++}$ & $CZ_{12}CCZ$ & $X$ basis & \textsf{ionq-11q} & 81\% \\
$\ket{+++}$ & $CZ_{13}CCZ$ & $X$ basis & \textsf{ionq-11q} & 74\% \\
$\ket{+++}$ & $CZ_{12}CZ_{13}CCZ$ & $X$ basis & \textsf{ionq-11q} & 84\% \\
$\ket{+++}$ & $CZ_{23}CCZ$ & $X$ basis & \textsf{ionq-11q} & 75\% \\
$\ket{+++}$ & $CZ_{12}CZ_{23}CCZ$ & $X$ basis & \textsf{ionq-11q} & 83\% \\
$\ket{+++}$ & $CZ_{13}CZ_{23}CCZ$ & $X$ basis & \textsf{ionq-11q} & 77\% \\
$\ket{+++}$ & $CZ_{12}CZ_{13}CZ_{23}CCZ$ & $X$ basis & \textsf{ionq-11q} & 77\% \\
$\ket{+++}$ & $I$ & $Z$ basis & \textsf{ionq-11q} & 73\% \\
$\ket{+++}$ & $CZ_{12}$ & $Z$ basis & \textsf{ionq-11q} & 78\% \\
$\ket{+++}$ & $CZ_{13}$ & $Z$ basis & \textsf{ionq-11q} & 71\% \\
$\ket{+++}$ & $CZ_{12}CZ_{13}$ & $Z$ basis & \textsf{ionq-11q} & 74\% \\
$\ket{+++}$ & $CZ_{23}$ & $Z$ basis & \textsf{ionq-11q} & 72\% \\
$\ket{+++}$ & $CZ_{12}CZ_{23}$ & $Z$ basis & \textsf{ionq-11q} & 72\% \\
$\ket{+++}$ & $CZ_{13}CZ_{23}$ & $Z$ basis & \textsf{ionq-11q} & 73\% \\
$\ket{+++}$ & $CZ_{12}CZ_{13}CZ_{23}$ & $Z$ basis & \textsf{ionq-11q} & 71\% \\
$\ket{+++}$ & $CCZ$ & $Z$ basis & \textsf{ionq-11q} & 74\% \\
$\ket{+++}$ & $CZ_{12}CCZ$ & $Z$ basis & \textsf{ionq-11q} & 74\% \\
$\ket{+++}$ & $CZ_{13}CCZ$ & $Z$ basis & \textsf{ionq-11q} & 70\% \\
$\ket{+++}$ & $CZ_{12}CZ_{13}CCZ$ & $Z$ basis & \textsf{ionq-11q} & 74\% \\
$\ket{+++}$ & $CZ_{23}CCZ$ & $Z$ basis & \textsf{ionq-11q} & 72\% \\
$\ket{+++}$ & $CZ_{12}CZ_{23}CCZ$ & $Z$ basis & \textsf{ionq-11q} & 73\% \\
$\ket{+++}$ & $CZ_{13}CZ_{23}CCZ$ & $Z$ basis & \textsf{ionq-11q} & 74\% \\
$\ket{+++}$ & $CZ_{12}CZ_{13}CZ_{23}CCZ$ & $Z$ basis & \textsf{ionq-11q} & 73\% \\
\hline\hline 
\end{tabular}
\end{center}
\caption{
Post-selection rates for the results shown in \cref{fig:X-+++,fig:Z-+++}, where the percentage indicates the proportion of shots that were accepted.
The average post-selection rate was 75\%.
}
\label{tab:ionq-+++}
\end{table}

\begin{table}[ht]
\begin{center}
\begin{tabular}{ c c c c c }
\hline\hline 
State & Gate & Measurement & Device & Post-selection rate \\
\hline
$\ket{\mathrm{GHZ}}$ & $I$ & $X$ basis & \textsf{ionq-11q} & 90\% \\
$\ket{\mathrm{GHZ}}$ & $CZ_{12}$ & $X$ basis & \textsf{ionq-11q} & 86\% \\
$\ket{\mathrm{GHZ}}$ & $CZ_{13}$ & $X$ basis & \textsf{ionq-11q} & 90\% \\
$\ket{\mathrm{GHZ}}$ & $CZ_{12}CZ_{13}$ & $X$ basis & \textsf{ionq-11q} & 78\% \\
$\ket{\mathrm{GHZ}}$ & $CZ_{23}$ & $X$ basis & \textsf{ionq-11q} & 83\% \\
$\ket{\mathrm{GHZ}}$ & $CZ_{12}CZ_{23}$ & $X$ basis & \textsf{ionq-11q} & 81\% \\
$\ket{\mathrm{GHZ}}$ & $CZ_{13}CZ_{23}$ & $X$ basis & \textsf{ionq-11q} & 82\% \\
$\ket{\mathrm{GHZ}}$ & $CZ_{12}CZ_{13}CZ_{23}$ & $X$ basis & \textsf{ionq-11q} & 79\% \\
$\ket{\mathrm{GHZ}}$ & $CCZ$ & $X$ basis & \textsf{ionq-11q} & 80\% \\
$\ket{\mathrm{GHZ}}$ & $CZ_{12}CCZ$ & $X$ basis & \textsf{ionq-11q} & 78\% \\
$\ket{\mathrm{GHZ}}$ & $CZ_{13}CCZ$ & $X$ basis & \textsf{ionq-11q} & 82\% \\
$\ket{\mathrm{GHZ}}$ & $CZ_{12}CZ_{13}CCZ$ & $X$ basis & \textsf{ionq-11q} & 87\% \\
$\ket{\mathrm{GHZ}}$ & $CZ_{23}CCZ$ & $X$ basis & \textsf{ionq-11q} & 83\% \\
$\ket{\mathrm{GHZ}}$ & $CZ_{12}CZ_{23}CCZ$ & $X$ basis & \textsf{ionq-11q} & 83\% \\
$\ket{\mathrm{GHZ}}$ & $CZ_{13}CZ_{23}CCZ$ & $X$ basis & \textsf{ionq-11q} & 82\% \\
$\ket{\mathrm{GHZ}}$ & $CZ_{12}CZ_{13}CZ_{23}CCZ$ & $X$ basis & \textsf{ionq-11q} & 84\% \\
$\ket{\mathrm{GHZ}}$ & $I$ & $Z$ basis & \textsf{ionq-11q} & 86\% \\
$\ket{\mathrm{GHZ}}$ & $CZ_{12}$ & $Z$ basis & \textsf{ionq-11q} & 81\% \\
$\ket{\mathrm{GHZ}}$ & $CZ_{13}$ & $Z$ basis & \textsf{ionq-11q} & 84\% \\
$\ket{\mathrm{GHZ}}$ & $CZ_{12}CZ_{13}$ & $Z$ basis & \textsf{ionq-11q} & 85\% \\
$\ket{\mathrm{GHZ}}$ & $CZ_{23}$ & $Z$ basis & \textsf{ionq-11q} & 81\% \\
$\ket{\mathrm{GHZ}}$ & $CZ_{12}CZ_{23}$ & $Z$ basis & \textsf{ionq-11q} & 90\% \\
$\ket{\mathrm{GHZ}}$ & $CZ_{13}CZ_{23}$ & $Z$ basis & \textsf{ionq-11q} & 84\% \\
$\ket{\mathrm{GHZ}}$ & $CZ_{12}CZ_{13}CZ_{23}$ & $Z$ basis & \textsf{ionq-11q} & 81\% \\
$\ket{\mathrm{GHZ}}$ & $CCZ$ & $Z$ basis & \textsf{ionq-11q} & 86\% \\
$\ket{\mathrm{GHZ}}$ & $CZ_{12}CCZ$ & $Z$ basis & \textsf{ionq-11q} & 80\% \\
$\ket{\mathrm{GHZ}}$ & $CZ_{13}CCZ$ & $Z$ basis & \textsf{ionq-11q} & 87\% \\
$\ket{\mathrm{GHZ}}$ & $CZ_{12}CZ_{13}CCZ$ & $Z$ basis & \textsf{ionq-11q} & 75\% \\
$\ket{\mathrm{GHZ}}$ & $CZ_{23}CCZ$ & $Z$ basis & \textsf{ionq-11q} & 74\% \\
$\ket{\mathrm{GHZ}}$ & $CZ_{12}CZ_{23}CCZ$ & $Z$ basis & \textsf{ionq-11q} & 83\% \\
$\ket{\mathrm{GHZ}}$ & $CZ_{13}CZ_{23}CCZ$ & $Z$ basis & \textsf{ionq-11q} & 75\% \\
$\ket{\mathrm{GHZ}}$ & $CZ_{12}CZ_{13}CZ_{23}CCZ$ & $Z$ basis & \textsf{ionq-11q} & 85\% \\
\hline\hline 
\end{tabular}
\end{center}
\caption{
Post-selection rates for the results shown in \cref{fig:X-GHZ,fig:Z-GHZ} for the IonQ device, where the percentage indicates the proportion of shots that were accepted.
The average post-selection rate was 83\%.
}
\label{tab:ionq-ghz}
\end{table}

\begin{table}[ht]
\begin{center}
\begin{tabular}{ c c c c c }
\hline\hline 
State & Gate & Measurement & Device & Post-selection rate \\
\hline
$\ket{\mathrm{GHZ}}$ & $I$ & $X$ basis & \textsf{ibmq\_mumbai} & 76\% \\
$\ket{\mathrm{GHZ}}$ & $CZ_{12}$ & $X$ basis & \textsf{ibmq\_mumbai} & 69\% \\
$\ket{\mathrm{GHZ}}$ & $CZ_{13}$ & $X$ basis & \textsf{ibmq\_mumbai} & 72\% \\
$\ket{\mathrm{GHZ}}$ & $CZ_{12}CZ_{13}$ & $X$ basis & \textsf{ibmq\_mumbai} & 79\% \\
$\ket{\mathrm{GHZ}}$ & $CZ_{23}$ & $X$ basis & \textsf{ibmq\_mumbai} & 69\% \\
$\ket{\mathrm{GHZ}}$ & $CZ_{12}CZ_{23}$ & $X$ basis & \textsf{ibmq\_mumbai} & 76\% \\
$\ket{\mathrm{GHZ}}$ & $CZ_{13}CZ_{23}$ & $X$ basis & \textsf{ibmq\_mumbai} & 78\% \\
$\ket{\mathrm{GHZ}}$ & $CZ_{12}CZ_{13}CZ_{23}$ & $X$ basis & \textsf{ibmq\_mumbai} & 80\% \\
$\ket{\mathrm{GHZ}}$ & $CCZ$ & $X$ basis & \textsf{ibmq\_mumbai} & 72\% \\
$\ket{\mathrm{GHZ}}$ & $CZ_{12}CCZ$ & $X$ basis & \textsf{ibmq\_mumbai} & 73\% \\
$\ket{\mathrm{GHZ}}$ & $CZ_{13}CCZ$ & $X$ basis & \textsf{ibmq\_mumbai} & 79\% \\
$\ket{\mathrm{GHZ}}$ & $CZ_{12}CZ_{13}CCZ$ & $X$ basis & \textsf{ibmq\_mumbai} & 76\% \\
$\ket{\mathrm{GHZ}}$ & $CZ_{23}CCZ$ & $X$ basis & \textsf{ibmq\_mumbai} & 73\% \\
$\ket{\mathrm{GHZ}}$ & $CZ_{12}CZ_{23}CCZ$ & $X$ basis & \textsf{ibmq\_mumbai} & 79\% \\
$\ket{\mathrm{GHZ}}$ & $CZ_{13}CZ_{23}CCZ$ & $X$ basis & \textsf{ibmq\_mumbai} & 79\% \\
$\ket{\mathrm{GHZ}}$ & $CZ_{12}CZ_{13}CZ_{23}CCZ$ & $X$ basis & \textsf{ibmq\_mumbai} & 72\% \\
$\ket{\mathrm{GHZ}}$ & $I$ & $Z$ basis & \textsf{ibmq\_mumbai} & 74\% \\
$\ket{\mathrm{GHZ}}$ & $CZ_{12}$ & $Z$ basis & \textsf{ibmq\_mumbai} & 70\% \\
$\ket{\mathrm{GHZ}}$ & $CZ_{13}$ & $Z$ basis & \textsf{ibmq\_mumbai} & 74\% \\
$\ket{\mathrm{GHZ}}$ & $CZ_{12}CZ_{13}$ & $Z$ basis & \textsf{ibmq\_mumbai} & 77\% \\
$\ket{\mathrm{GHZ}}$ & $CZ_{23}$ & $Z$ basis & \textsf{ibmq\_mumbai} & 71\% \\
$\ket{\mathrm{GHZ}}$ & $CZ_{12}CZ_{23}$ & $Z$ basis & \textsf{ibmq\_mumbai} & 71\% \\
$\ket{\mathrm{GHZ}}$ & $CZ_{13}CZ_{23}$ & $Z$ basis & \textsf{ibmq\_mumbai} & 75\% \\
$\ket{\mathrm{GHZ}}$ & $CZ_{12}CZ_{13}CZ_{23}$ & $Z$ basis & \textsf{ibmq\_mumbai} & 73\% \\
$\ket{\mathrm{GHZ}}$ & $CCZ$ & $Z$ basis & \textsf{ibmq\_mumbai} & 76\% \\
$\ket{\mathrm{GHZ}}$ & $CZ_{12}CCZ$ & $Z$ basis & \textsf{ibmq\_mumbai} & 78\% \\
$\ket{\mathrm{GHZ}}$ & $CZ_{13}CCZ$ & $Z$ basis & \textsf{ibmq\_mumbai} & 78\% \\
$\ket{\mathrm{GHZ}}$ & $CZ_{12}CZ_{13}CCZ$ & $Z$ basis & \textsf{ibmq\_mumbai} & 77\% \\
$\ket{\mathrm{GHZ}}$ & $CZ_{23}CCZ$ & $Z$ basis & \textsf{ibmq\_mumbai} & 78\% \\
$\ket{\mathrm{GHZ}}$ & $CZ_{12}CZ_{23}CCZ$ & $Z$ basis & \textsf{ibmq\_mumbai} & 77\% \\
$\ket{\mathrm{GHZ}}$ & $CZ_{13}CZ_{23}CCZ$ & $Z$ basis & \textsf{ibmq\_mumbai} & 77\% \\
$\ket{\mathrm{GHZ}}$ & $CZ_{12}CZ_{13}CZ_{23}CCZ$ & $Z$ basis & \textsf{ibmq\_mumbai} & 71\% \\
\hline\hline 
\end{tabular}
\end{center}
\caption{
Post-selection rates for the results shown in \cref{fig:X-GHZ,fig:Z-GHZ} for the IBM device, where the percentage indicates the proportion of shots that were accepted.
The average post-selection rate was 75\%.
}
\label{tab:ibm-+++}
\end{table}

\clearpage

\section{Additional hardware details}
\label{app:char}

\subsection{IBM}

The layout of the \textsf{ibmq\_mumbai} device is shown in~\cref{fig:ibm-conn}.
The demonstrations were carried out on 14/04/23.
The 1-qubit and 2-qubit gate characterization data provided by IBM for this date are given in~\cref{tab:ibm-1q,tab:ibm-2q}.
The qubit characterization data provided by IBM for this date are given in~\cref{tab:ibm-q-prop} and the qubit reset time was $\SI{3612}{\nano\second}$ for all qubits.

\begin{figure}[ht]
    \centering
    \begin{tikzpicture}[scale=0.3]
        \node[draw,black,circle] (0) at (0,0){0};
        \node[draw,black,circle] (1) at (4,0){1};
        \node[draw,black,circle] (4) at (8,0){4};
        \node[draw,black,circle] (7) at (12,0){7};
        \node[draw,black,circle] (10) at (16,0){10};
        \node[draw,black,circle] (12) at (20,0){12};
        \node[draw,black,circle] (15) at (24,0){15};
        \node[draw,black,circle] (18) at (28,0){18};
        \node[draw,black,circle] (21) at (32,0){21};
        \node[draw,black,circle] (23) at (36,0){23};
        \node[draw,black,circle] (6) at (12,4){6};
        \node[draw,black,circle] (17) at (28,4){17};
        \node[draw,black,circle] (2) at (4,-4){2};
        \node[draw,black,circle] (3) at (4,-8){3};
        \node[draw,black,circle] (13) at (20,-4){13};
        \node[draw,black,circle] (14) at (20,-8){14};
        \node[draw,black,circle] (24) at (36,-4){24};
        \node[draw,black,circle] (25) at (36,-8){25};
        \node[draw,black,circle] (5) at (8,-8){5};
        \node[draw,black,circle] (8) at (12,-8){8};
        \node[draw,black,circle] (11) at (16,-8){11};
        \node[draw,black,circle] (16) at (24,-8){16};
        \node[draw,black,circle] (19) at (28,-8){19};
        \node[draw,black,circle] (22) at (32,-8){22};
        \node[draw,black,circle] (26) at (40,-8){26};
        \node[draw,black,circle] (9) at (12,-12){9};
        \node[draw,black,circle] (20) at (28,-12){20};
        \draw (0) -- (1);
        \draw (1) -- (4);
        \draw (4) -- (7);
        \draw (7) -- (10);
        \draw (10) -- (12);
        \draw (12) -- (15);
        \draw (15) -- (18);
        \draw (18) -- (21);
        \draw (21) -- (23);
        \draw (6) -- (7);
        \draw (17) -- (18);
        \draw (1) -- (2);
        \draw (2) -- (3);
        \draw (12) -- (13);
        \draw (13) -- (14);
        \draw (23) -- (24);
        \draw (24) -- (25);
        \draw (3) -- (5);
        \draw (5) -- (8);
        \draw (8) -- (11);
        \draw (11) -- (14);
        \draw (14) -- (16);
        \draw (16) -- (19);
        \draw (19) -- (22);
        \draw (22) -- (25);
        \draw (25) -- (26);
        \draw (8) -- (9);
        \draw (19) -- (20);
    \end{tikzpicture}
    \caption{
   Layout of the \textsf{ibmq\_mumbai} device.
   Vertices represent qubits and edges represent the availability of entangling gates between the two endpoints.
   The mapping from the qubits of the \NKD{8}{3}{2} code to the qubits of the device was $(q_0, 7)$, $(q_1, 4)$, $(q_2, 1)$, $(q_3, 10)$, $(q_4, 2)$, $(q_5, 1{\color{red}2})$, $(q_6, 13)$, $(q_7, 3)$.
    }
    \label{fig:ibm-conn}
\end{figure}

\begin{table}[ht]
\begin{minipage}{0.45\textwidth}
\centering
        \begin{tabular}{ c c c c }
            \hline\hline            
            Qubit & Gate & Error & Time \\
            \hline
            0 & $I,X,\sqrt{X}$ & 0.0002951 & $\SI{35.56}{\nano\second}$ \\
            1 & $I,X,\sqrt{X}$ & 0.0001877 & $\SI{35.56}{\nano\second}$ \\
            2 & $I,X,\sqrt{X}$ & 0.0001816 & $\SI{35.56}{\nano\second}$ \\
            3 & $I,X,\sqrt{X}$ & 0.0002697 & $\SI{35.56}{\nano\second}$ \\
            4 & $I,X,\sqrt{X}$ & 0.0003707 & $\SI{35.56}{\nano\second}$ \\
            5 & $I,X,\sqrt{X}$ & 0.0002801 & $\SI{35.56}{\nano\second}$ \\
            6 & $I,X,\sqrt{X}$ & 0.000267 & $\SI{35.56}{\nano\second}$ \\
            7 & $I,X,\sqrt{X}$ & 0.0001768 & $\SI{35.56}{\nano\second}$ \\
            8 & $I,X,\sqrt{X}$ & 0.0001585 & $\SI{35.56}{\nano\second}$ \\
            9 & $I,X,\sqrt{X}$ & 0.0003164 & $\SI{35.56}{\nano\second}$ \\
            10 & $I,X,\sqrt{X}$ & 0.0002868 & $\SI{35.56}{\nano\second}$ \\
            11 & $I,X,\sqrt{X}$ & 0.0002381 & $\SI{35.56}{\nano\second}$ \\
            12 & $I,X,\sqrt{X}$ & 0.0001866 & $\SI{35.56}{\nano\second}$ \\
            13 & $I,X,\sqrt{X}$ & 0.0001609 & $\SI{35.56}{\nano\second}$ \\
            14 & $I,X,\sqrt{X}$ & 0.000182 & $\SI{35.56}{\nano\second}$ \\
            15 & $I,X,\sqrt{X}$ & 0.0004084 & $\SI{35.56}{\nano\second}$ \\
            16 & $I,X,\sqrt{X}$ & 0.0001818 & $\SI{35.56}{\nano\second}$ \\
            17 & $I,X,\sqrt{X}$ & 0.002198 & $\SI{35.56}{\nano\second}$ \\
            18 & $I,X,\sqrt{X}$ & 0.0002243 & $\SI{35.56}{\nano\second}$ \\
            19 & $I,X,\sqrt{X}$ & 0.0001976 & $\SI{35.56}{\nano\second}$ \\
            20 & $I,X,\sqrt{X}$ & 0.0002066 & $\SI{35.56}{\nano\second}$ \\
            21 & $I,X,\sqrt{X}$ & 0.0004652 & $\SI{35.56}{\nano\second}$ \\
            22 & $I,X,\sqrt{X}$ & 0.000166 & $\SI{35.56}{\nano\second}$ \\
            23 & $I,X,\sqrt{X}$ & 0.0003324 & $\SI{35.56}{\nano\second}$ \\
            24 & $I,X,\sqrt{X}$ & 0.0001602 & $\SI{35.56}{\nano\second}$ \\
            25 & $I,X,\sqrt{X}$ & 0.0002232 & $\SI{35.56}{\nano\second}$ \\
            26 & $I,X,\sqrt{X}$ & 0.0002714 & $\SI{35.56}{\nano\second}$ \\
            \hline\hline
        \end{tabular}
    \caption{
        1-qubit gate characterization data for \textsf{ibmq\_mumbai} on 14/04/23.
    }
    \label{tab:ibm-1q}
\end{minipage}
\begin{minipage}{0.45\textwidth}
\centering
\begin{tabular}{ c c c c }
            \hline\hline            
            Qubits & Gate & Error & Time \\
            \hline
            3,2 & CNOT & 0.008789 & $\SI{433.8}{\nano\second}$ \\
            2,3 & CNOT & 0.008789 & $\SI{469.3}{\nano\second}$ \\
            14,11 & CNOT & 0.005348 & $\SI{391.1}{\nano\second}$ \\
            11,14 & CNOT & 0.005348 & $\SI{426.7}{\nano\second}$ \\
            5,8 & CNOT & 0.01016 & $\SI{604.4}{\nano\second}$ \\
            8,5 & CNOT & 0.01016 & $\SI{640.0}{\nano\second}$ \\
            12,13 & CNOT & 0.00516 & $\SI{547.6}{\nano\second}$ \\
            13,12 & CNOT & 0.00516 & $\SI{583.1}{\nano\second}$ \\
            13,14 & CNOT & 0.004559 & $\SI{320.0}{\nano\second}$ \\
            14,13 & CNOT & 0.004559 & $\SI{355.6}{\nano\second}$ \\
            22,19 & CNOT & 0.004825 & $\SI{327.1}{\nano\second}$ \\
            19,22 & CNOT & 0.004825 & $\SI{362.7}{\nano\second}$ \\
            3,5 & CNOT & 0.01117 & $\SI{476.4}{\nano\second}$ \\
            5,3 & CNOT & 0.01117 & $\SI{512.0}{\nano\second}$ \\
            18,21 & CNOT & 0.01269 & $\SI{497.8}{\nano\second}$ \\
            21,18 & CNOT & 0.01269 & $\SI{533.3}{\nano\second}$ \\
            25,22 & CNOT & 0.005982 & $\SI{448.0}{\nano\second}$ \\
            22,25 & CNOT & 0.005982 & $\SI{483.6}{\nano\second}$ \\
            2,1 & CNOT & 0.01161 & $\SI{704.0}{\nano\second}$ \\
            1,2 & CNOT & 0.01161 & $\SI{739.6}{\nano\second}$ \\
            8,11 & CNOT & 0.01248 & $\SI{604.4}{\nano\second}$ \\
            11,8 & CNOT & 0.01248 & $\SI{640.0}{\nano\second}$ \\
            10,12 & CNOT & 0.007643 & $\SI{604.4}{\nano\second}$ \\
            12,10 & CNOT & 0.007643 & $\SI{640.0}{\nano\second}$ \\
            10,7 & CNOT & 0.007437 & $\SI{398.2}{\nano\second}$ \\
            7,10 & CNOT & 0.007437 & $\SI{433.8}{\nano\second}$ \\
            20,19 & CNOT & 0.005657 & $\SI{369.8}{\nano\second}$ \\
            19,20 & CNOT & 0.005657 & $\SI{405.3}{\nano\second}$ \\
            23,21 & CNOT & 0.008342 & $\SI{362.7}{\nano\second}$ \\
            21,23 & CNOT & 0.008342 & $\SI{398.2}{\nano\second}$ \\
            6,7 & CNOT & 0.007037 & $\SI{248.9}{\nano\second}$ \\
            7,6 & CNOT & 0.007037 & $\SI{284.4}{\nano\second}$ \\
            17,18 & CNOT & 0.01205 & $\SI{248.9}{\nano\second}$ \\
            18,17 & CNOT & 0.01205 & $\SI{284.4}{\nano\second}$ \\
            4,7 & CNOT & 0.009587 & $\SI{604.4}{\nano\second}$ \\
            7,4 & CNOT & 0.009587 & $\SI{640.0}{\nano\second}$ \\
            8,9 & CNOT & 0.008003 & $\SI{604.4}{\nano\second}$ \\
            9,8 & CNOT & 0.008003 & $\SI{640.0}{\nano\second}$ \\
            16,19 & CNOT & 0.01609 & $\SI{682.7}{\nano\second}$ \\
            19,16 & CNOT & 0.01609 & $\SI{718.2}{\nano\second}$ \\
            24,23 & CNOT & 0.01297 & $\SI{604.4}{\nano\second}$ \\
            23,24 & CNOT & 0.01297 & $\SI{640.0}{\nano\second}$ \\
            4,1 & CNOT & 0.006135 & $\SI{312.9}{\nano\second}$ \\
            1,4 & CNOT & 0.006135 & $\SI{348.4}{\nano\second}$ \\
            15,18 & CNOT & 0.006877 & $\SI{305.8}{\nano\second}$ \\
            18,15 & CNOT & 0.006877 & $\SI{341.3}{\nano\second}$ \\
            16,14 & CNOT & 0.005754 & $\SI{291.6}{\nano\second}$ \\
            14,16 & CNOT & 0.005754 & $\SI{327.1}{\nano\second}$ \\
            26,25 & CNOT & 0.006879 & $\SI{312.9}{\nano\second}$ \\
            25,26 & CNOT & 0.006879 & $\SI{348.4}{\nano\second}$ \\
            0,1 & CNOT & 0.007626 & $\SI{419.6}{\nano\second}$ \\
            1,0 & CNOT & 0.007626 & $\SI{455.1}{\nano\second}$ \\
            15,12 & CNOT & 0.006138 & $\SI{369.8}{\nano\second}$ \\
            12,15 & CNOT & 0.006138 & $\SI{405.3}{\nano\second}$ \\
            24,25 & CNOT & 0.007386 & $\SI{433.8}{\nano\second}$ \\
            25,24 & CNOT & 0.007386 & $\SI{469.3}{\nano\second}$ \\
            \hline\hline
        \end{tabular}
        \caption{
        2-qubit gate characterization data for \textsf{ibmq\_mumbai} on 14/04/23.
    }
    \label{tab:ibm-2q}
\end{minipage}
\end{table}

\begin{table}[ht]
    \begin{center}
        \begin{tabular}{ c c c c c c c c c}
            \hline\hline
            Qubit & $T_1$ & $T_2$ & Frequency & Anharmonicity & Readout error & $p(0|1)$ & $p(1|0)$ & Readout time \\ 
            \hline
            0 & $\SI{100.5}{\micro\second}$ & $\SI{125.9}{\micro\second}$ & $\SI{5.071}{\giga\hertz}$ & $\SI{-0.3285}{\giga\hertz}$ & 0.0435 & 0.0714 & 0.0156 & $\SI{3576}{\nano\second}$ \\
            1 & $\SI{116.4}{\micro\second}$ & $\SI{163.8}{\micro\second}$ & $\SI{4.93}{\giga\hertz}$ & $\SI{-0.3313}{\giga\hertz}$ & 0.0401 & 0.0476 & 0.0326 & $\SI{3576}{\nano\second}$ \\
            2 & $\SI{94.52}{\micro\second}$ & $\SI{114.8}{\micro\second}$ & $\SI{4.67}{\giga\hertz}$ & $\SI{-0.3369}{\giga\hertz}$ & 0.0153 & 0.0206 & 0.01 & $\SI{3576}{\nano\second}$ \\
            3 & $\SI{103.2}{\micro\second}$ & $\SI{168.3}{\micro\second}$ & $\SI{4.889}{\giga\hertz}$ & $\SI{-0.3312}{\giga\hertz}$ & 0.057 & 0.0622 & 0.0518 & $\SI{3576}{\nano\second}$ \\
            4 & $\SI{105.9}{\micro\second}$ & $\SI{47.41}{\micro\second}$ & $\SI{5.021}{\giga\hertz}$ & $\SI{-0.3302}{\giga\hertz}$ & 0.03 & 0.0424 & 0.0176 & $\SI{3576}{\nano\second}$ \\
            5 & $\SI{86.9}{\micro\second}$ & $\SI{151.6}{\micro\second}$ & $\SI{4.969}{\giga\hertz}$ & $\SI{-0.3298}{\giga\hertz}$ & 0.1603 & 0.178 & 0.1426 & $\SI{3576}{\nano\second}$ \\
            6 & $\SI{122.5}{\micro\second}$ & $\SI{81.42}{\micro\second}$ & $\SI{4.966}{\giga\hertz}$ & $\SI{-0.3293}{\giga\hertz}$ & 0.0169 & 0.0236 & 0.0102 & $\SI{3576}{\nano\second}$ \\
            7 & $\SI{125.2}{\micro\second}$ & $\SI{164.3}{\micro\second}$ & $\SI{4.894}{\giga\hertz}$ & $\SI{-0.331}{\giga\hertz}$ & 0.0162 & 0.0232 & 0.0092 & $\SI{3576}{\nano\second}$ \\
            8 & $\SI{212.2}{\micro\second}$ & $\SI{256.6}{\micro\second}$ & $\SI{4.792}{\giga\hertz}$ & $\SI{-0.3326}{\giga\hertz}$ & 0.018 & 0.0192 & 0.0168 & $\SI{3576}{\nano\second}$ \\
            9 & $\SI{149.8}{\micro\second}$ & $\SI{119.8}{\micro\second}$ & $\SI{4.955}{\giga\hertz}$ & $\SI{-0.3306}{\giga\hertz}$ & 0.0119 & 0.0184 & 0.0054 & $\SI{3576}{\nano\second}$ \\
            10 & $\SI{99.87}{\micro\second}$ & $\SI{115.1}{\micro\second}$ & $\SI{4.959}{\giga\hertz}$ & $\SI{-0.3309}{\giga\hertz}$ & 0.0211 & 0.0288 & 0.0134 & $\SI{3576}{\nano\second}$ \\
            11 & $\SI{148.1}{\micro\second}$ & $\SI{137.1}{\micro\second}$ & $\SI{4.666}{\giga\hertz}$ & $\SI{-0.3326}{\giga\hertz}$ & 0.032 & 0.0422 & 0.0218 & $\SI{3576}{\nano\second}$ \\
            12 & $\SI{151.6}{\micro\second}$ & $\SI{249.3}{\micro\second}$ & $\SI{4.743}{\giga\hertz}$ & $\SI{-0.333}{\giga\hertz}$ & 0.0348 & 0.0414 & 0.0282 & $\SI{3576}{\nano\second}$ \\
            13 & $\SI{192.4}{\micro\second}$ & $\SI{236.7}{\micro\second}$ & $\SI{4.889}{\giga\hertz}$ & $\SI{-0.3281}{\giga\hertz}$ & 0.0124 & 0.0184 & 0.0064 & $\SI{3576}{\nano\second}$ \\
            14 & $\SI{156.7}{\micro\second}$ & $\SI{223.5}{\micro\second}$ & $\SI{4.78}{\giga\hertz}$ & $\SI{-0.3326}{\giga\hertz}$ & 0.0214 & 0.0276 & 0.0152 & $\SI{3576}{\nano\second}$ \\
            15 & $\SI{111.2}{\micro\second}$ & $\SI{36.64}{\micro\second}$ & $\SI{4.858}{\giga\hertz}$ & $\SI{-0.3332}{\giga\hertz}$ & 0.024 & 0.0364 & 0.0116 & $\SI{3576}{\nano\second}$ \\
            16 & $\SI{131.2}{\micro\second}$ & $\SI{187.7}{\micro\second}$ & $\SI{4.98}{\giga\hertz}$ & $\SI{-0.3299}{\giga\hertz}$ & 0.0634 & 0.075 & 0.0518 & $\SI{3576}{\nano\second}$ \\
            17 & $\SI{63.8}{\micro\second}$ & $\SI{114.2}{\micro\second}$ & $\SI{5.003}{\giga\hertz}$ & $\SI{-0.3299}{\giga\hertz}$ & 0.0206 & 0.0314 & 0.0098 & $\SI{3576}{\nano\second}$ \\
            18 & $\SI{132.7}{\micro\second}$ & $\SI{172.1}{\micro\second}$ & $\SI{4.781}{\giga\hertz}$ & $\SI{-0.3331}{\giga\hertz}$ & 0.0787 & 0.0852 & 0.0722 & $\SI{3576}{\nano\second}$ \\
            19 & $\SI{169.2}{\micro\second}$ & $\SI{209.2}{\micro\second}$ & $\SI{4.81}{\giga\hertz}$ & $\SI{-0.3321}{\giga\hertz}$ & 0.0306 & 0.0316 & 0.0296 & $\SI{3576}{\nano\second}$ \\ 
            20 & $\SI{96.24}{\micro\second}$ & $\SI{218.2}{\micro\second}$ & $\SI{5.048}{\giga\hertz}$ & $\SI{-0.328}{\giga\hertz}$ & 0.0137 & 0.0226 & 0.0048 & $\SI{3576}{\nano\second}$ \\ 
            21 & $\SI{105.7}{\micro\second}$ & $\SI{162.3}{\micro\second}$ & $\SI{4.943}{\giga\hertz}$ & $\SI{-0.3313}{\giga\hertz}$ & 0.0418 & 0.0666 & 0.017 & $\SI{3576}{\nano\second}$ \\ 
            22 & $\SI{161.2}{\micro\second}$ & $\SI{194.7}{\micro\second}$ & $\SI{4.911}{\giga\hertz}$ & $\SI{-0.3318}{\giga\hertz}$ & 0.0145 & 0.0232 & 0.0058 & $\SI{3576}{\nano\second}$ \\ 
            23 & $\SI{102.2}{\micro\second}$ & $\SI{166.8}{\micro\second}$ & $\SI{4.893}{\giga\hertz}$ & $\SI{-0.3315}{\giga\hertz}$ & 0.0471 & 0.0582 & 0.036 & $\SI{3576}{\nano\second}$ \\ 
            24 & $\SI{139.5}{\micro\second}$ & $\SI{41.14}{\micro\second}$ & $\SI{4.671}{\giga\hertz}$ & $\SI{-0.3359}{\giga\hertz}$ & 0.0152 & 0.0226 & 0.0078 & $\SI{3576}{\nano\second}$ \\ 
            25 & $\SI{146.9}{\micro\second}$ & $\SI{182.3}{\micro\second}$ & $\SI{4.759}{\giga\hertz}$ & $\SI{-0.3336}{\giga\hertz}$ & 0.0159 & 0.0184 & 0.0134 & $\SI{3576}{\nano\second}$ \\ 
            26 & $\SI{135.8}{\micro\second}$ & $\SI{244.6}{\micro\second}$ & $\SI{4.954}{\giga\hertz}$ & $\SI{-0.3295}{\giga\hertz}$ & 0.018 & 0.0214 & 0.0146 & $\SI{3576}{\nano\second}$ \\ 
            \hline\hline
        \end{tabular}
    \end{center}
    \caption{
        Qubit characterization data for \textsf{ibmq\_mumbai} on 14/04/23.
        $p(i|j)$ is the probability of measuring a $\ket i$ state, given that a $\ket j$ state was prepared.
    }
    \label{tab:ibm-q-prop}
\end{table}

\subsection{IonQ}

The \textsf{ionq-11q} device has 11 qubits and all-to-all connectivity.
The characterization data provided by IonQ for the dates that the demonstrations were carried out are given in~\cref{tab:ionq-char}.

\begin{table}[ht]
    \begin{center}
        \begin{tabular}{ c c c c c c c c c c }
            \hline\hline 
            Date & 1q gate fidelity & 2q gate fidelity & SPAM fidelity & $T_1$ & $T_2$ & 1q gate time & 2q gate time & Readout time & Reset time \\
            \hline
            26/07/23 & $0.9958$ & $0.9652$ & $0.99752$ & $\SI{10000}{\second}$ & $\SI{0.2}{\second}$ & $\SI{10}{\micro\second}$ & $\SI{200}{\micro\second}$ & $\SI{130}{\micro\second}$ & $\SI{20}{\micro\second}$ \\
            15/08/23 & 0.9976 & 0.9906 & 0.99752 & $\SI{10000}{\second}$ & $\SI{0.2}{\second}$ & $\SI{10}{\micro\second}$ & $\SI{200}{\micro\second}$ & $\SI{130}{\micro\second}$ & $\SI{20}{\micro\second}$ \\
            \hline\hline
        \end{tabular}
    \end{center}
    \caption{
        Characterization data for \textsf{ionq-11q}.
        The demonstrations shown in~\cref{fig:X-+++,fig:Z-+++} were conducted on 26/07/23 and the demonstrations shown in~\cref{fig:X-GHZ,fig:Z-GHZ} were conducted on 17/08/23 (the data for 15/08/23 shown in the table is the nearest in time to when the relevant demonstrations were carried out).
        The fidelities are mean values.
        SPAM stands for state-preparation and measurement.
    }
    \label{tab:ionq-char}
\end{table}

\end{document}